%% file: main.tex
\documentclass[lettersize,journal]{IEEEtran}

\usepackage{cite}
\usepackage[T1]{fontenc}
\usepackage{amsmath,amssymb,amsfonts}
\usepackage{graphicx}
\usepackage{textcomp}
\usepackage{xcolor}

\usepackage[caption=false,font=footnotesize]{subfig}
\usepackage[bookmarks=false]{hyperref}
\usepackage{tikz}
\usetikzlibrary{shapes,arrows}
\usetikzlibrary{positioning,calc}
\usepackage{pgfplots}
\pgfplotsset{compat=1.18}
\usepackage{booktabs}
\usepackage{url}
\usepackage{orcidlink}

\usepackage[linesnumbered, ruled, vlined]{algorithm2e}

\begin{document}

\title{Priority-Based Network Access with Strict-Priority Scheduling in PLCA Multidrop Networks}

\author{Mohamed Elattar \orcidlink{0009-0009-9283-3301}, Zhe Lou \orcidlink{0009-0009-0305-9167},  Georg Carle \orcidlink{0000-0002-2347-1839}

\thanks{Mohamed Elattar is with the Applied Network Technology Lab, Huawei Heisenberg Research Center, Munich, Germany, and also with the Chair of Network Architectures and Services, Technical University of Munich (TUM), Munich, Germany (e-mail: mohamed.osama.elattar@huawei.com; m.elattar@tum.de).}

\thanks{Zhe Lou is with the Applied Network Technology Lab, Huawei Heisenberg Research Center, Munich, Germany (e-mail: zhe.lou@huawei.com).}

\thanks{Georg Carle is with the Chair of Network Architectures and Services, Technical University of Munich, Munich, Germany (e-mail: carle@net.in.tum.de).}

\thanks{This work has been submitted to the IEEE for possible publication. Copyright may be transferred without notice, after which this version may no longer be accessible.}

}

\maketitle
\input{Abstract}

\begin{IEEEkeywords}
Physical Layer Collision Avoidance (PLCA), Priority-Based Network Access, Strict Priority Scheduling, Traffic Differentiation, Bounded Access Latency, Shared-medium Ethernet, 10BASE-T1S/T1M

\end{IEEEkeywords}

\input{Ch1_Introduction}
\input{Ch2_Background}
\input{Ch3_RelatedWork}

\input{Ch4_systemModel}

\input{Ch5_PLCASP}
\input{Ch6_derivations}
\input{Ch7_evaluation}

\input{Ch8_conclusion}

\bibliographystyle{IEEEtran}
\bibliography{IEEEabrv,reference}

\end{document}

%% file: Abstract.tex
\begin{abstract}

Shared-medium Ethernet is gaining renewed attention across industrial domains, driven by the adoption of 10BASE-T1S. In such multidrop networks, collision-free and deterministic medium access is achieved through Physical Layer Collision Avoidance (PLCA). However, PLCA follows a node-based cyclic access structure that does not account for traffic priority, leading to high access latency for time-sensitive traffic and limited scalability with increasing network size. Existing PLCA extensions improve latency for selected devices but remain restricted to node-based access and do not support network-wide traffic differentiation. To address these limitations, this paper introduces a priority-based network access and scheduling mechanism for multidrop networks. The proposed approach decouples medium access from node-based ordering and enables class-based prioritization, allowing high-priority traffic from any device to be transmitted ahead of lower-priority traffic. It further supports a configuration that emulates Strict Priority (SP) scheduling, providing low and bounded access latency for highest-priority traffic. Analytical bounds on the worst-case access latency are derived,  validated, and shown to be tight. Performance evaluation demonstrates that the proposed mechanism reduces the maximum observed access latency for high-priority traffic by over 65\% compared to PLCA, with latency increasing by less than 1\% as the network grows from 8 to 40 devices.
\end{abstract}

%% file: Ch1_Introduction.tex
\section{Introduction}
\label{Sec: Introduction}

\IEEEPARstart{S}{hared-medium} Ethernet is re-emerging across diverse domains, including industrial automation, robotics, and in-vehicle networks. In automotive systems, Ethernet has become the dominant in-vehicle communication technology, driven by the transition toward zonal architectures \cite{automotive-architectures}. Similarly, industrial networks are transitioning from heterogeneous, task-specific fieldbus systems to unified Ethernet-based infrastructures, eliminating the need for dedicated gateways and enabling seamless integration of edge computing and virtualized control \cite{SPE2025}.

In such multidrop shared-medium environments, collision-free and deterministic medium access is essential to support latency-sensitive and safety-critical applications \cite{8021DG2025}. Physical Layer Collision Avoidance (PLCA), standardized in 10BASE-T1S \cite{8023cg}\cite{8023da}, addresses these requirements by assigning exclusive transmission opportunities to devices, thereby eliminating collisions and providing bounded access latency.

Despite these advantages, PLCA exhibits fundamental limitations. Medium access is strictly ordered according to node identifiers (IDs), without considering traffic priority. As a result, latency-sensitive traffic may experience significant delays while waiting for transmissions from lower-priority traffic. Moreover, due to its cyclic access structure, the worst-case access latency of PLCA grows linearly with the number of devices, limiting its scalability in large networks. The IEEE 802.3 PLCA Enhancements Study Group is exploring enhancements to address these limitations, including support for differentiated services and improvements in latency performance and scalability~\cite{PLCAEnhancementsSG}.

Several extensions have been proposed to introduce prioritization and differentiated service into PLCA networks. Some approaches prioritize selected devices by modifying their access order~\cite{PLCANP,PriorityBasedPLCA} or transmission frequency~\cite{PLCAMIDs,WeightedMultiID,PLCAMIDsppt}. Others regulate transmissions according to packet priorities or flow timing requirements~\cite{PriorityAdvertisingPLCA,APPshaper,FS2,CBSPLCA,TASPLCA}. While these approaches improve service differentiation, they still retain device-associated transmission opportunities. Consequently, medium access for high-priority traffic remains affected by the access opportunities assigned to its transmitting device, limiting the ability to provide priority access independently of which device generates the traffic.

This paper analyzes the access latency of PLCA and its variant, PLCA with node prioritization (PLCA-NP), highlighting the impact of node-based access on latency and scalability. To address these limitations, we introduce a priority-based network access and scheduling mechanism for shared-medium multidrop networks. The proposed mechanism assigns transmission opportunities to traffic classes rather than individual devices. Any device with pending traffic of the corresponding class can participate in its transmission opportunity, with contention resolved through a deterministic ID-based tree mechanism. This enables network-wide traffic differentiation, allowing high-priority traffic from any device to gain access ahead of pending lower-priority traffic. Furthermore, a configuration that emulates Strict Priority (SP) scheduling is introduced, providing low and bounded access latency for the highest-priority traffic.

This paper substantially extends our previous work in \cite{PLCA_SP} as follows. First, we provide a formal worst-case latency analysis of the resolution mechanism. Second, building on these results, we formally derive the resulting worst-case access latency, covering the single-packet assumption as well as the general multi-packet case and the worst-case queueing delay. Third, we validate the derived bounds through an exhaustive search over all possible resolution scenarios, confirming that the bounds are both valid and tight. Fourth, we provide a comprehensive performance evaluation across a wider range of network sizes and traffic loads. Finally, we refine the proposed mechanism along two aspects: we introduce two strategies for handling interrupted resolution processes, an aspect not discussed in \cite{PLCA_SP}, and we retain the BEACON message from standard PLCA to ensure robust synchronization for devices joining or rejoining during an ongoing cycle.

The remainder of this paper is organized as follows. Section~\ref{Sec: Background} provides background on shared-medium Ethernet and PLCA. Section~\ref{Sec: Related Work} reviews existing PLCA extensions. Section~\ref{Sec: System Model} presents the system model and analyzes the access latency of PLCA and PLCA-NP. Section~\ref{Sec: Proposed Mechanism} introduces the proposed mechanism. Section~\ref{sec:Worst-Case Access Latency Analysis under SP Configuration} derives its worst-case latency bounds and Section~\ref{Sec: Evaluation} evaluates its performance. Finally, Section~\ref{Sec: Conclusion} concludes the paper.

%% file: Ch2_Background.tex
\section{Background} 
\label{Sec: Background}

This section reviews network architectures and medium access mechanisms, with emphasis on shared-medium environments. It then discusses random access and collision resolution mechanisms, and introduces deterministic medium access through PLCA.

\subsection{Network Architectures and Medium Access}
\label{Subsec: Network Architectures and Medium Access}
Communication networks can broadly be categorized into switched point-to-point architectures and shared-medium (multidrop) architectures. In point-to-point, each device is connected to a central switch via a dedicated physical link, creating independent communication channels. This eliminates contention between devices and allows multiple transmissions to occur simultaneously. 

In such networks, access is controlled by the switch, which enables the straightforward implementation of scheduling mechanisms. These range from legacy disciplines such as Strict Priority (SP) and fair queueing~\cite{WRR,DRR,GPS_scheduling}, to advanced schedulers such as the Urgency-Based Scheduler (UBS)~\cite{UBS} and the Paternoster scheduler~\cite{Paternoster}, as well as Time-Sensitive Networking (TSN) mechanisms~\cite{8021Q2022}, including Time-Aware Shaping (TAS) and Multiple Cyclic Queuing and Forwarding (Multi-CQF)~\cite{Multi-cqf-pdf}. The availability of dedicated resources and centralized control allows the scheduler to deterministically select packets based on their priority, thereby facilitating Quality of Service (QoS) guarantees, including bounded latency.

In contrast, shared-medium networks connect multiple devices to a common communication channel, forming a single collision domain. Since all devices share the same medium, simultaneous transmissions result in collisions and unsuccessful reception. Consequently, medium access must be coordinated among distributed devices to ensure that, at any given time, at most one device transmits. 

The absence of centralized control and dedicated communication resources makes the realization of scheduling policies in shared-medium networks fundamentally more challenging. While priority-based scheduling mechanisms such as SP are straightforward to implement in switched architectures, achieving equivalent behavior in a distributed shared-medium environment requires explicit coordination of transmission attempts across devices. This challenge becomes especially pronounced when aiming to provide low and bounded latency guarantees.

\subsection{Random Access and Collision Resolution}
\label{Subsec: Random Access and Collision Resolution}

In shared-medium Ethernet, medium access is typically governed by Carrier Sense Multiple Access with Collision Detection (CSMA/CD) \cite{CSMACD}, in which distributed devices sense the channel and defer transmission when ongoing activity is detected. Despite carrier sensing, simultaneous transmission attempts may still occur. To resolve such contention, CSMA/CD employs a collision resolution mechanism, namely exponential backoff, where devices involved in a collision delay their retransmissions by selecting a random waiting interval. 

Tree-based algorithms provide a more structured resolution approach by recursively partitioning colliding devices into subsets corresponding to branches of a logical tree \cite{TreeAlgorithm}. Devices are randomly assigned to subsets, and transmission opportunities are granted to each subset sequentially. Variants such as the modified tree algorithm \cite{MTA} and the dynamic tree algorithm \cite{DTA} further improve resolution efficiency by avoiding foreseen collisions and adapting the number of subsets to the number of contending devices, respectively. More recently, tree algorithms have also been combined with successive interference cancellation to further reduce resolution latency~\cite{SIC2007}\cite{SIC2024}.

Despite these enhancements, CSMA/CD and resolution mechanisms (both back-off-based and tree-based), rely on stochastic behavior, as transmission attempts and resolution decisions involve randomization. Consequently, the resulting access delay is variable and cannot be deterministically bounded. This limitation makes random access schemes unsuitable for time-sensitive applications requiring strict latency guarantees.

\subsection{PLCA: Physical Layer Collision Avoidance}

Physical Layer Collision Avoidance (PLCA), standardized in IEEE 802.3, enables deterministic medium access in shared-medium Ethernet networks such as 10BASE-T1S~\cite{8023cg} and 10BASE-T1M~\cite{8023da}. It eliminates collisions entirely among devices sharing the same communication medium. The core principle of PLCA is to allocate exclusive Transmission Opportunities (TOs) to devices in a cyclic manner, arranged according to their unique node identifiers. Within each PLCA cycle, initiated by a BEACON message, devices are granted TOs sequentially to either transmit pending data or yield through a short idle interval.

Although PLCA provides bounded worst-case access latency, it exhibits fundamental limitations. Due to its cyclic, node-based access structure, the worst-case access latency grows linearly with the number of devices, significantly limiting scalability in large networks. Furthermore, PLCA does not account for traffic priority, as transmission order is determined solely by node identifiers. Consequently, latency-sensitive traffic may experience significant delays due to preceding lower-priority transmissions.

Several extensions have been proposed to address these limitations by introducing prioritized access, which reduces access latency in PLCA networks. These approaches are reviewed in Section~\ref{Sec: Related Work}.

%% file: Ch3_RelatedWork.tex
\section{Related Work} 
\label{Sec: Related Work}

Existing work has explored several approaches to improve latency and provide differentiated service in PLCA networks. We first review methods that adapt device-associated transmission opportunities by modifying access order,  transmission frequency, or cycle timing. We then discuss approaches that regulate transmissions according to packet priorities or flow timing requirements.

\subsection{Node-Based Access Allocation and Prioritization}
PLCA with Node Prioritization (PLCA-NP), introduced in \cite{PLCANP, priority_plca}, designates a subset of devices as priority nodes, each assigned a priority rank and a corresponding request slot within the inter-packet gap (IPG). These devices can request access after an ongoing transmission completes, without waiting for their regular TOs. The regular PLCA sequence resumes when no priority node requests access. In addition, \cite{PLCANP} combines node prioritization with packet segmentation to reduce the blocking time caused by long ongoing transmissions.

A Multi-ID approach, presented in \cite{PLCAMIDs}, assigns multiple node identifiers to selected devices, granting them multiple TOs within each PLCA cycle. Additional identifiers are allocated according to the latency requirements of the generated flows, with the corresponding TOs distributed across the cycle to reduce waiting times. The weighted Multi-ID method in \cite{WeightedMultiID} considers both deadline urgency and offered traffic load when allocating identifiers to balance the access requirements of latency-sensitive and heavily loaded devices.

The proposal in \cite{PLCAMIDsppt} controls transmission eligibility across PLCA cycles. Each device is assigned a cycle interval and an offset, and may transmit during its TO only when a modulo condition on the cycle counter is satisfied. Devices with tighter latency requirements can therefore transmit every cycle, while others are eligible less frequently.

Priority-based transmit ordering is proposed in \cite{PriorityBasedPLCA} by assigning earlier identifiers to higher-priority nodes and allowing them to issue a priority request before a lower-priority TO is used. The request prompts the coordinator to transmit a new BEACON, restarting the access sequence from the highest-priority opportunities without waiting for the full cycle to complete. Periodic PLCA, introduced in \cite{Periodic_PLCA}, addresses latency variation by enforcing a fixed cycle duration through periodic BEACON transmission.

Despite their effectiveness in reducing access latency or latency variation, these approaches retain node-based access allocation, while adapting access ordering, transmission frequency, or cycle timing. Consequently, preferential access depends on the opportunities assigned to each device, limiting the ability to prioritize time-sensitive traffic independently of its originating device.

\subsection{Traffic-Based Access Control and Scheduling}

The proposal in \cite{PriorityAdvertisingPLCA} introduces a priority-advertisement phase before the data phase of each PLCA cycle. Devices advertise the priorities of their pending packets and use the exchanged information to determine which packets may be transmitted under the selected scheduling policy. This enables priority-aware access across devices while retaining node-associated TOs, at the cost of an additional signaling phase in each cycle.

PLCA with Interrupt, proposed in \cite{InterruptPLCA}, distinguishes express packets from ordinary packets. While ordinary packets follow the regular PLCA TOs, express packets can interrupt ongoing lower-priority transmissions. Once an express transmission completes, the interrupted device regains access and retransmits its packet.

Other approaches provide differentiated service to traffic flows generated by the same device. In \cite{APPshaper}, an application-level shaping mechanism prioritizes packets from the flow with the shortest relative deadline. The Frame Segmentation and Shaping approach in \cite{FS2} further controls the size of transmitted packets through segmentation and regulates each flow's transmission rate through shaping. These parameters are configured according to the flows’ timing requirements to meet their relative deadlines.

The interaction between traffic shaping and PLCA medium access has also been studied analytically. The work in \cite{CBSPLCA} develops a network-calculus model for Credit-Based Shaping (CBS) over PLCA, using a credit-update policy that freezes credit when the device has no TO and during higher-priority control transmissions. Similarly, \cite{TASPLCA} develops a network-calculus model for integrating Time-Aware Shaping (TAS) with PLCA. The analysis accounts for the interaction between fixed-duration TAS schedules and variable-duration PLCA cycles to derive deterministic delay bounds for scheduled traffic.

Although these approaches \cite{PriorityAdvertisingPLCA,APPshaper,FS2,CBSPLCA,TASPLCA} enable traffic differentiation, they retain conventional node-based TOs. Consequently, even a packet given higher priority must wait for the transmission opportunity assigned to its device, so its access latency remains affected by the node-based access order.

%% file: Ch4_systemModel.tex
\section{System Model and Access Latency Analysis of PLCA and PLCA-NP} 
\label{Sec: System Model}

This section establishes the system model and analyzes the access latency characteristics of standard PLCA and its extension, PLCA-NP. Among existing extensions, PLCA-NP achieves very low access latency for designated priority nodes through static prioritization, making it a suitable benchmark for evaluating latency performance.

\subsection{Network and Traffic Model}
\label{subsec:system model}

We consider a shared-medium (multidrop) Ethernet network consisting of $N$ devices connected to a common communication channel with capacity $C$. All devices share a single collision domain and operate in half-duplex mode, such that at most one device may transmit at any given time to ensure successful reception.

Medium access is governed by the PLCA mechanism. PLCA operates as a reconciliation sublayer between the Medium Access Control (MAC) and physical layers, leveraging CSMA/CD functionalities, such as carrier sensing, to prevent physical collisions. 

Each device in the PLCA domain is assigned a unique node identifier $n~\in~\{1,\dots,N\}$, which determines its position in the access order. Devices may generate traffic belonging to one or more priority classes. We assume a total of $K$ priority classes, indexed by $k \in \{1,\dots,K\}$, where a lower index corresponds to a higher priority (i.e., $k=1$ denotes the highest-priority traffic). The notation used throughout the paper is summarized in Table~\ref{tab:notation}.

In the following, we focus on access latency, defined as the time elapsed between a packet becoming ready for transmission at a device and the instant its transmission begins. We investigate medium-access behavior and latency under an idealized model with synchronized devices, no channel-induced errors, and neglected propagation and processing delays. These effects and physical layer realization are not addressed in this paper and are left for future work.

\begin{table}[t]
\centering
\caption{Summary of Notation}
\label{tab:notation}
\begin{tabular}{l p{5.8cm}}
\toprule
\textbf{Symbol} & \textbf{Description} \\
\midrule
$\mathcal{N}$ & Set of all devices in the system \\
$N = |\mathcal{N}|$ & Number of devices \\
$n \in \{1,\dots,N\}$ & Device (node) identifier \\
$\hat{n}$ & Transmission Opportunity (TO) counter \\
\midrule
$K$ & Number of traffic classes \\
$k \in \{1,\dots,K\}$ & Class index (lower index indicates higher priority) \\
$\hat{k}$ & Class Transmission Opportunity (CTO) counter \\

\midrule
$\mathcal{M}$ & Set of contending devices \\
$M = |\mathcal{M}|$ & Number of contending devices \\
$\mathcal{S}_m$ & $m$-th group in the contention resolution tree \\
\midrule
$T_\mathrm{b}$ & BEACON message duration \\
$T_\mathrm{c}$ & COMMIT message (collision interval) duration \\
$T_\mathrm{s}$ & Silent interval (idle interval) duration  \\
$T_{\mathrm{pkt,}\max}$ & Maximum packet size transmission duration\\
\midrule
$x \in \{\mathrm{FT}, \mathrm{LS}, \mathrm{FS}\}$ & Resolution traversal scenario: FT, LS, FS \\
$I_{\max}^{(x)}$ & Maximum number of idle intervals in scenario $x$ \\
$C_{\max}^{(x)}$ & Maximum number of collisions in scenario $x$ \\
$I_{\mathrm{C,}\max}^{(x)}$ & Maximum number of idle intervals conditioned on maximum collisions in scenario $x$ \\
$C_{\mathrm{I,}\max}^{(x)}$ & Maximum number of collisions conditioned on maximum idle intervals in scenario $x$ \\

\bottomrule
\end{tabular}
\end{table}

\subsection{PLCA Operation and Access Latency Analysis}
\label{subsec: PLCA Operation and Access Latency Analysis}

PLCA provides collision-free medium access by assigning exclusive TOs to devices in a cyclic order. Each device maintains local state variables required for distributed coordination, including the total number of devices ($N$), a transmission opportunity counter ($\hat{n}$), and a TO timer that defines the silent interval ($T_\mathrm{s}$). 

At the beginning of each PLCA cycle, the coordinator (typically the device with $n=1$) transmits a BEACON message, synchronizing all devices and resetting their TO counters to $\hat{n}~=~1$. The TO counter is then incremented sequentially across all devices. When the counter matches a device’s identifier ($\hat{n} = n$), that device is granted exclusive access to the medium.

If the device has pending data, it transmits a COMMIT message to reserve the channel, followed by the data packet. Otherwise, it remains silent, yielding its TO. After either transmission or silent interval of a duration~$T_\mathrm{s}$, the TO counter advances to the next device. Once the last TO ($\hat{n} = N$) completes, a new BEACON message is issued to start the next cycle.

The resulting transmission behavior is illustrated in Fig.~\ref{subfig: Transmission behavior of PLCA}. The duration of a PLCA cycle depends on device activity: it is minimized when all devices yield their TOs and maximized when all devices transmit packets of maximum size.

In the worst case, a packet arriving immediately after a device’s TO has elapsed must wait for a complete cycle before transmission (Fig. \ref{subfig: Worst-case access latency in an 8-device PLCA network}). The corresponding worst-case access latency is given by
\begin{equation}
\label{equ: PLCA maximum latency}
L_{n\mathrm{,}\max}^{\text{PLCA}} = T_\mathrm{b} + N T_\mathrm{c} + (N-1)  T_{\mathrm{pkt,}\max},
\end{equation}
where $T_\mathrm{b}$ denotes the BEACON message duration, $T_\mathrm{c}$ the COMMIT message duration, and $T_{\mathrm{pkt,}\max}$ the transmission time of the maximum packet size.

As evident from (\ref{equ: PLCA maximum latency}), the worst-case access latency in PLCA grows linearly with the number of devices. For instance, in a network with $N=8$ devices operating at 10~Mbps, the worst-case latency reaches $8.5068$~ms (see Table~\ref{tab:PLCA Parameters} for configuration parameters). Since PLCA does not support traffic differentiation and the access order is strictly determined by node identifiers, latency-sensitive traffic is subject to the same delay. These limitations restrict the applicability of PLCA in large-scale networks and for time-sensitive applications. For example, time-critical control traffic in in-vehicle networks is commonly reported to require access latencies on the order of~1~ms~\cite{cakir2019qos}\cite{peng2023survey}.

\input{Fig1_PLCA_Trasmission_Sequence}

\begin{table}[t]
\centering
\renewcommand{\arraystretch}{1.3}
\setlength{\tabcolsep}{15pt}
\caption{Configuration Parameters (\(C = 10 \text{\upshape{ Mbps}} \)) }
\label{tab:PLCA Parameters}
\begin{tabular}{c | c c}
\toprule
\textbf{Parameter} & \textbf{Length} & \textbf{Duration} \\ 
\midrule
BEACON Message& \(20\text{ bits}\) & \(2 \text{ } \mu \text{s}\) \\ 
COMMIT Message& \(5\text{ bits}\) & \(0.5 \text{ } \mu \text{s}\) \\ 
Silent/Idle Interval & \( - \) & \(3.2 \text{ } \mu \text{s}\) \\ 
Maximum Packet Size & \(1518\text{ byte}\) & \(1214.4 \text{ } \mu \text{s}\) \\ 
\bottomrule
\end{tabular}
\end{table}

\subsection{PLCA-NP Operation and Access Latency Analysis}
\label{Subsec: PLCA-NP Operation and Access Latency Analysis}

PLCA-NP extends the standard PLCA mechanism by enabling a predefined subset of devices, referred to as priority nodes, to obtain faster access to the shared medium. The primary objective of PLCA-NP is to reduce access latency for traffic generated by these selected devices. 

In PLCA-NP, all devices continue to participate in the standard PLCA cycle as described in Section~\ref{subsec: PLCA Operation and Access Latency Analysis}. In addition, priority nodes are given the opportunity to request medium access during the IPG following the completion of any data transmission. To support this, the IPG is partitioned into $K$ request slots, each slot associated with one priority class.

Each priority class $k$, where $k < K$, is uniquely mapped to a single priority node. That is, at most one device in the network may transmit traffic of priority class $k$, while all remaining devices operate as best-effort nodes and are assigned to the lowest-priority class ($k=K$).

A priority node of class $k$ may request expedited access by asserting its signal during its designated request slot within the IPG. If granted, the node transmits a COMMIT message followed by its data packet. If no priority node issues a request, the PLCA cycle resumes, granting access to the next device in the cyclic order.

Fig.~\ref{subfig: Trasmission sequence of PLCA-NP with two priority nodes} illustrates the transmission behavior of PLCA-NP for $K=3$ priority classes, where devices~7 and~4 are configured as priority nodes. Device~7, corresponding to the highest-priority class ($k=1$), bypasses the cyclic TO order and transmits immediately after device~2 completes its transmission. Device~4 ($k=2$) can access the medium only if the highest-priority node (device~7) does not request access. The standard PLCA cycle resumes only when all priority nodes refrain from requesting expedited transmission.

\input{Fig2_PLCA_NP_Trasmission_Sequence}

In the worst case, the access latency for the highest-priority node ($k=1$), illustrated in Fig.~\ref{subfig: Guaranteed maximum access latency for the highest-priority node using PLCA-NP}, is given by
\begin{equation}
\label{equ: PLCA-NP maximum latency}
        L_{k=1\mathrm{,}\max}^{\text{PLCA-NP}} = T_\mathrm{b} + (N-2) T_\mathrm{s} + T_\mathrm{c} + T_{\mathrm{pkt,}\max} + T_{\mathrm{IPG}},
\end{equation}
where $T_\mathrm{IPG}$ denotes the IPG duration, which for a network with $K$ priority classes is obtained as 
\begin{equation}
\label{equ: PLCA-NP IPG}
    T_{\mathrm{IPG}} = K T_\mathrm{s} .
\end{equation}
For example, in a network with $N=8$ devices and $K=3$ priority classes, the worst-case access latency for the highest priority node is $1.2457$~ms (Table~\ref{tab:PLCA Parameters}).

While PLCA-NP effectively reduces access latency for designated priority nodes, this improvement is limited to those specific devices. It assumes that latency-critical traffic originates only from a predefined subset of devices and therefore cannot provide latency guarantees when such traffic is generated by arbitrary nodes.

Moreover, PLCA-NP preserves the underlying cyclic access structure of PLCA. As a result, the access latency remains dependent on the network size $N$ and additionally increases with the number of supported priority nodes $K$, as evident from (\ref{equ: PLCA-NP maximum latency}) and (\ref{equ: PLCA-NP IPG}).

%% file: Fig1_PLCA_Trasmission_Sequence.tex
\input{DefineColors}

\input{DefineStyles_P3}

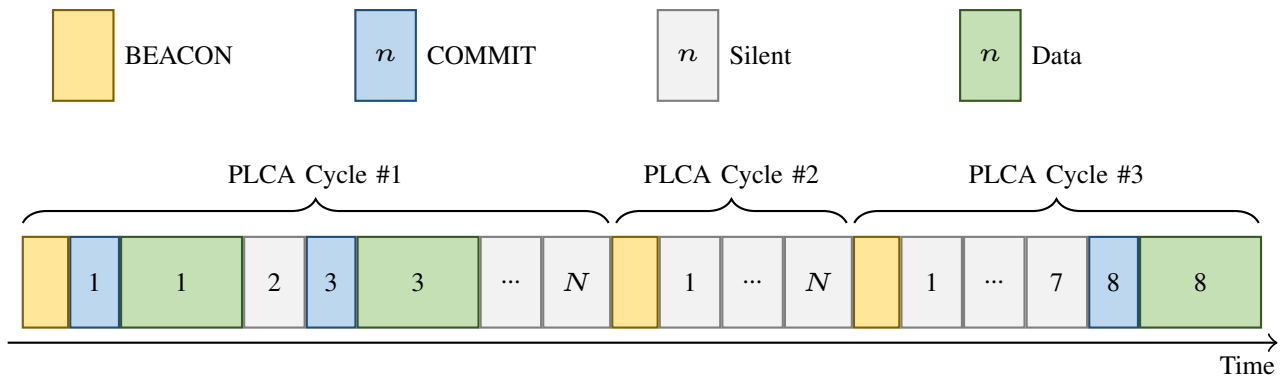
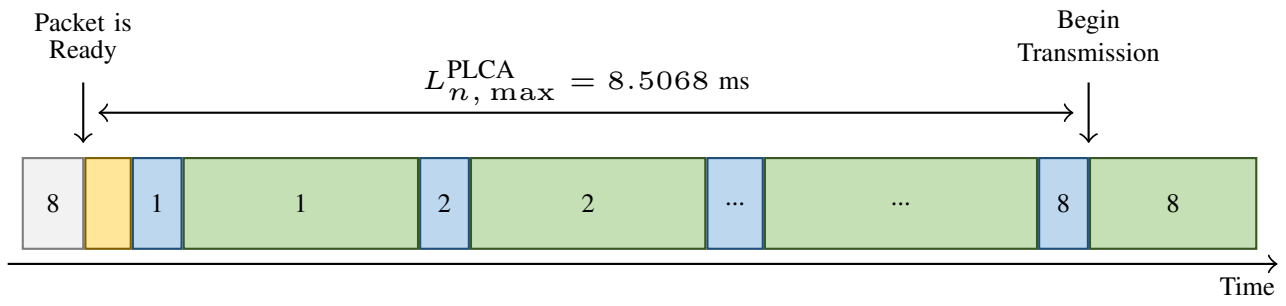
\begin{figure}[!t]
    \centering

    \subfloat[Transmission sequence of PLCA.]{%
        \begin{tikzpicture}

        \begin{scope}[shift={(0, 1.9)}] 
        
        \node () [beaconStd] at (0.5,0) {};
        \node[right] at (0.65, 0) {\tiny BEACON};

        \node () [commitStd] at (2.5,0) {\tiny $n$};
        \node[right] at (2.65, 0) {\tiny COMMIT};

        \node () [silentStd] at (4.5,0) {\tiny $n$};
        \node[right] at (4.65, 0) {\tiny Silent};

        \node () [dataStd] at (6.5,0) {\tiny $n$};
        \node[right] at (6.65, 0) {\tiny Data};
        \end{scope}

        \draw[ ->] (0,0) -- (8.4,0) node[right]{};
        \node at (8.2, -0.15) {\tiny Time}; 

        \draw[decorate, decoration={brace, amplitude=1ex, raise=-1ex}] (0.1,0.95)  -- ++ (3.875,0);
        \node [] at (0.1 + 1.9375, 0.95 + 0.15) {\tiny PLCA Cycle \#1};

        \draw[decorate, decoration={brace, amplitude=1ex, raise=-1ex}] (4.025,0.95)  -- ++ (1.525,0);
        \node [] at (4.025 + 0.7625, 0.95 + 0.15) {\tiny PLCA Cycle \#2};

        \draw[decorate, decoration={brace, amplitude=1ex, raise=-1ex}] (5.6,0.95)  -- ++ (2.675,0);
        \node [] at (5.6 + 1.3375, 0.95 + 0.15) {\tiny PLCA Cycle \#3};

        \node (b10) [beacon] at (0.25,0.4) {};
        
        \node (c11) [commit, right=0cm of b10] {\tiny 1};
        \node (p11) [min_packet, right=0cm of c11] {\tiny 1};
        
        \node (s12) [silent, right=0cm of p11] {\tiny 2};
        
        \node (c13) [commit, right=0cm of s12] {\tiny 3};
        \node (p13) [min_packet, right=0cm of c13] {\tiny 3};

        \node (s14) [silent, right=0cm of p13] {\tiny ...};

        \node (s1n) [silent, right=0cm of s14] {\tiny $N$};

        \node (b20) [beacon, right=0cm of s1n] {};
        \node (s21) [silent, right=0cm of b20] {\tiny 1};
        \node (s22) [silent, right=0cm of s21] {\tiny ...};
        \node (s25) [silent, right=0cm of s22] {\tiny $N$};

        \node (b30) [beacon, right=0cm of s25] {};
        \node (s31) [silent, right=0cm of b30] {\tiny 1};
        \node (s32) [silent, right=0cm of s31] {\tiny ...};
        \node (s37) [silent, right=0cm of s32] {\tiny 7};
        \node (c38) [commit, right=0cm of s37] {\tiny 8};
        \node (p31) [min_packet, right=0cm of c38] {\tiny 8};

        \end{tikzpicture}
        \label{subfig: Transmission behavior of PLCA}
    }

   \subfloat[Worst-case access latency in PLCA.]{%
        \begin{tikzpicture}
            
        \draw[->] (0,0) -- (8.4,0) node[right]{};
        
        \node at (8.2, -0.15) {\tiny Time}; 

        \node (s18) [silent] at (0.3,0.4) {\tiny 8};
        \node (b10) [beacon, right=0cm of s18] {};
        \node (c11) [commit, right=0cm of b10] {\tiny 1};
        \node (p11) [max_packet, right=0cm of c11] {\tiny 1};
        \node (c22) [commit, right=0cm of p11] {\tiny 2};
        \node (p22) [max_packet, right=0cm of c22] {\tiny 2};
        \node (c2s) [commit, right=0cm of p22] {\tiny ...};
        \node (p2s) [inf_packet, right=0cm of c2s] {\tiny ...};
        \node (c2n) [commit, right=0cm of p2s] {\tiny 8};
        \node (p2n) [ave_packet, right=0cm of c2n] {\tiny 8};

        \draw[->] (0.5,1.2) -- (0.5,0.8); 
        \draw[->] (7.15,1.2) -- (7.15,0.8);
        \draw[<->] (0.5 + 0.1,1.0) -- (7.15 - 0.1,1.0);
        \node[] at (0.5, 1.2 + 0.2 + 0.2) {\tiny Packet is};
        \node[] at (0.5, 1.2 + 0.2) {\tiny Ready};
        \node[] at (0.5 + 3.325, 1.0 + 0.2) {\tiny $L_{n\mathrm{,}\max}^{\text{PLCA}} = 8.5068$ ms};
        \node[] at (7.15, 1.2 + 0.2 + 0.2) {\tiny Begin};
        \node[] at (7.15, 1.2 + 0.2) {\tiny Transmission};

        \end{tikzpicture}
        \label{subfig: Worst-case access latency in an 8-device PLCA network}
    }
    
    \caption{Operation and worst-case access latency of PLCA in an $N=8$ device network.}
    \label{fig: PLCA operation and access latency in an 8-device network}

\end{figure}

%% file: DefineColors.tex
\definecolor{GrayBlueOLL}{HTML}{44546A}
\definecolor{GrayBlueOLM}{HTML}{333f50}
\definecolor{GrayBlueOLD}{HTML}{222A35}

\definecolor{GrayBlueFillL}{HTML}{D6DCE5}
\definecolor{GrayBlueFillD}{HTML}{ADB9CA}

\definecolor{BlueOLL}{HTML}{5B9BD5}
\definecolor{BlueOLM}{HTML}{2E75B6}
\definecolor{BlueOLD}{HTML}{1F4E79}

\definecolor{BlueFillL}{HTML}{DEEBF7}
\definecolor{BlueFillD}{HTML}{BDD7EE}

\definecolor{GreenOLL}{HTML}{70AD47}
\definecolor{GreenOLM}{HTML}{548235}
\definecolor{GreenOLD}{HTML}{385723}

\definecolor{GreenFillL}{HTML}{E2F0D9}
\definecolor{GreenFillD}{HTML}{C5E0B4}

\definecolor{WhiteOLL}{HTML}{BFBFBF}
\definecolor{WhiteOLM}{HTML}{A6A6A6}
\definecolor{WhiteOLD}{HTML}{7F7F7F}

\definecolor{WhiteFillL}{HTML}{F2F2F2}
\definecolor{WhiteFillD}{HTML}{D9D9D9}

\definecolor{GrayOLL}{HTML}{A5A5A5}
\definecolor{GrayOLM}{HTML}{7C7C7C}
\definecolor{GrayOLD}{HTML}{525252}

\definecolor{GrayFillL}{HTML}{EDEDED}
\definecolor{GrayFillD}{HTML}{DBDBDB}

\definecolor{YellowOLL}{HTML}{FFC000}
\definecolor{YellowOLM}{HTML}{BF9000}
\definecolor{YellowOLD}{HTML}{7F6000}

\definecolor{YellowFillL}{HTML}{FFF2CC}
\definecolor{YellowFillD}{HTML}{FFE699}

\definecolor{OrangeOLL}{HTML}{ED7D31}
\definecolor{OrangeOLM}{HTML}{C55A11}
\definecolor{OrangeOLD}{HTML}{843C0C}

\definecolor{OrangeFillL}{HTML}{FBE5D6}
\definecolor{OrangeFillD}{HTML}{F8CBAD}

\definecolor{PriorityOL}{HTML}{C0392B} 
\definecolor{PriorityFill}{HTML}{F5B7B1}

\definecolor{CollisionOL}{HTML}{16A085} 
\definecolor{CollisionFill}{HTML}{A3E4D7}

%% file: DefineStyles_P3.tex
\input{DefineColors}
\tikzstyle{beacon} = [rectangle, draw=YellowOLD, fill=YellowFillD, minimum height=0.6cm, minimum width=0.3cm]

\tikzstyle{commit} = [rectangle, draw=BlueOLD, fill=BlueFillD, minimum height=0.6cm, minimum width=0.2cm]

\tikzstyle{ipg_commit} = [rectangle, draw=BlueOLD, fill=BlueFillD, minimum height=0.5cm, minimum width=0.1cm]

\tikzstyle{CTO_commit} = [rectangle, draw=BlueOLD, fill=BlueFillD, minimum height=0.45cm, minimum width=0.1cm]

\tikzstyle{silent} = [rectangle, draw=WhiteOLD, fill=WhiteFillL, minimum height=0.6cm, minimum width=0.4cm]

\tikzstyle{ipg_silent} = [rectangle, draw=GrayOLD, fill=GrayFillD, minimum height=0.5cm, minimum width=0.3 cm]

\tikzstyle{CTO_silent} = [rectangle, draw=WhiteOLD, fill=WhiteFillL, minimum height=0.45cm, minimum width=0.3cm]

\tikzstyle{min_packet} = [rectangle, draw=GreenOLD, fill=GreenFillD, minimum   height=0.6cm, minimum width=0.8cm]

\tikzstyle{xsmall_packet} = [rectangle, draw=GreenOLD, fill=GreenFillD, minimum   height=0.6cm, minimum width=0.9cm]

\tikzstyle{small_packet} = [rectangle, draw=GreenOLD, fill=GreenFillD, minimum   height=0.6cm, minimum width=1cm]

\tikzstyle{ave_packet} = [rectangle, draw=GreenOLD, fill=GreenFillD, minimum height=0.6cm, minimum width=1.1cm]

\tikzstyle{large_packet} = [rectangle, draw=GreenOLD, fill=GreenFillD, minimum height=0.6cm, minimum width=1.2cm]

\tikzstyle{xlarge_packet} = [rectangle, draw=GreenOLD, fill=GreenFillD, minimum height=0.6cm, minimum width=1.3cm]

\tikzstyle{xxlarge_packet} = [rectangle, draw=GreenOLD, fill=GreenFillD, minimum height=0.6cm, minimum width=1.4cm]

\tikzstyle{xxxlarge_packet} = [rectangle, draw=GreenOLD, fill=GreenFillD, minimum height=0.6cm, minimum width=1.5cm]

\tikzstyle{max_packet} = [rectangle, draw=GreenOLD, fill=GreenFillD, minimum height=0.6cm, minimum width=1.55cm]

\tikzstyle{inf_packet} = [rectangle, draw=GreenOLD, fill=GreenFillD, minimum height=0.6cm, minimum width=1.8cm]

\tikzstyle{ipg_priority} = [rectangle, draw=PriorityOL, fill=PriorityFill, minimum height=0.5cm, minimum width=0.1cm]

\tikzstyle{ipg1} = [rectangle, draw=OrangeOLD, fill=OrangeFillD, minimum height=0.6cm, minimum width=0.82cm]

\tikzstyle{ipg2} = [rectangle, draw=OrangeOLD, fill=OrangeFillD, minimum height=0.6cm, minimum width=1.22cm]

\tikzstyle{ipg3} = [rectangle, draw=OrangeOLD, fill=OrangeFillD, minimum height=0.6cm, minimum width=1.22cm]

\tikzstyle{priority} = [rectangle, draw=PriorityOL, fill=PriorityFill, minimum height=0.6cm, minimum width=0.6cm]

\tikzstyle{priority2} = [rectangle, draw=PriorityOL, fill=PriorityFill, minimum height=0.6cm, minimum width=0.4cm]

\tikzstyle{collision} = [rectangle, draw=GrayBlueOLD, fill=GrayBlueFillD, minimum height=0.6cm, minimum width=0.2cm]

\tikzstyle{pkt} = [rectangle, draw=GreenOLD, fill=GreenFillD, minimum height=0.45cm, minimum width=0.8cm]

\tikzstyle{beaconStd} = [rectangle, draw=YellowOLD, fill=YellowFillD, minimum height=0.6cm, minimum width=0.4cm]

\tikzstyle{commitStd} = [rectangle, draw=BlueOLD, fill=BlueFillD, minimum height=0.6cm, minimum width=0.4cm]

\tikzstyle{silentStd} = [rectangle, draw=WhiteOLD, fill=WhiteFillL, minimum height=0.6cm, minimum width=0.4cm]

\tikzstyle{dataStd} = [rectangle, draw=GreenOLD, fill=GreenFillD, minimum   height=0.6cm, minimum width=0.4cm]

\tikzstyle{ipgStd} = [rectangle, draw=OrangeOLD, fill=OrangeFillD, minimum height=0.6cm, minimum width=0.4cm]

\tikzstyle{ipg_silentStd} = [rectangle, draw=GrayOLD, fill=GrayFillD, minimum height=0.6cm, minimum width=0.4cm]

\tikzstyle{ipg_priorityStd} = [rectangle, draw=PriorityOL, fill=PriorityFill, minimum height=0.6cm, minimum width=0.4cm]

%% file: Fig2_PLCA_NP_Trasmission_Sequence.tex
\input{DefineStyles_P3}
\begin{figure}[!t]
    \centering
    \subfloat[Transmission sequence of PLCA-NP with two priority nodes ($K=3$).]{    
        \begin{tikzpicture}

        \begin{scope}[shift={(0, 1.9)}] 
        
        \node () [ipgStd] at (0.5,0) {};
        \node[right] at (0.65, 0) {\tiny Inter-Packet Gap (IPG)};

        \node () [ipg_priorityStd] at (3.75,0) {\tiny $k$};
        \node[right] at (3.9, 0) {\tiny Priority Request};

        \node () [ipg_silentStd] at (6.5,0) {\tiny $k$};
        \node[right] at (6.65, 0) {\tiny Priority Silent};
        \end{scope}

        \draw[ ->] (0,0) -- (8.4,0) node[right]{};
        
        \node at (8.2, -0.15) {\tiny Time}; 

        \draw[decorate, decoration={brace, amplitude=1ex, raise=-1ex}] (1.975,0.95)  -- ++ (0.8,0);
        \node [] at (1.975 + 0.4, 0.95 + 0.15) {\tiny IPG};

        \draw[decorate, decoration={brace, amplitude=1ex, raise=-1ex}] (3.725,0.95)  -- ++ (1.2,0);
        \node [] at (3.725 + 0.6, 0.95 + 0.15) {\tiny IPG};

        \draw[decorate, decoration={brace, amplitude=1ex, raise=-1ex}] (5.875,0.95)  -- ++ (1.2,0);
        \node [] at (5.875 + 0.6, 0.95 + 0.15) {\tiny IPG};

        \node (b20) [beacon] at (0.25,0.4) {};
        \node (s21) [silent, right=0cm of b20 ] {\tiny 1};
        \node (s22) [commit, right=0cm of s21] {\tiny 2};
        \node (p22) [min_packet, right=0cm of s22] {\tiny 2};
        
        \node (ipg21) [ipg1, right=0cm of p22] {\tiny };
        \node (pr21) [ipg_priority, right=0.05cm of p22] {\tiny 1};
        \node (c21) [ipg_commit, right=0.05cm of pr21] {\tiny 7};
        \node (p22) [xsmall_packet, right=0cm of ipg21] {\tiny 7};
        
        \node (ipg22) [ipg2, right=0cm of p22] {\tiny };
        \node (pr22) [ipg_silent, right=0.05cm of p22] {\tiny 1};
        \node (pr22) [ipg_priority, right=0.05cm of pr22] {\tiny 2};
        \node (c22) [ipg_commit, right=0.05cm of pr22] {\tiny 4};
        \node (p23) [xsmall_packet, right=0cm of ipg22] {\tiny 4};
        
        \node (ipg23) [ipg3, right=0cm of p23] {\tiny };
        \node (pr23) [ipg_silent, right=0.05cm of p23] {\tiny 1};
        \node (pr23) [ipg_silent, right=0.05cm of pr23] {\tiny 2};
        \node (c23) [ipg_commit, right=0.05cm of pr23] {\tiny 3};
        \node (p24) [ave_packet, right=0cm of ipg23] {\tiny 3};

        \end{tikzpicture}
        
        \label{subfig: Trasmission sequence of PLCA-NP with two priority nodes}
    }

   \subfloat[Worst-case access latency for the highest-priority node in PLCA-NP.]{%
        \begin{tikzpicture}
           
        \draw[ ->] (0,0) -- (8.4,0) node[right]{};

        \node at (8.2, -0.15) {\tiny Time};

        \node (s16) [silent] at (0.3,0.4) {\tiny 7};
        \node (s17) [silent, right=0cm of s16] {\tiny 8};

        \node (b20) [beacon, right=0cm of s17] {};

        \node (s21) [silent, right=0cm of b20] {\tiny 1};
        \node (s22) [silent, right=0cm of s21] {\tiny ...};
        \node (s25) [silent, right=0cm of s22] {\tiny 5};
        \node (s26) [commit, right=0cm of s25] {\tiny 6};
        \node (p26) [max_packet, right=0cm of s26] {\tiny 6};

        \node (ipg26) [ipg1, right=0cm of p26] {\tiny };
        \node (pr26) [ipg_priority, right=0.05cm of p26] {\tiny 1};
        \node (c26) [ipg_commit, right=0.05cm of pr26] {\tiny 7};
        \node (p26) [xsmall_packet, right=0cm of ipg26] {\tiny 7};

        \node (ipg27) [ipg2, right=0cm of p26] {\tiny };
        \node (pr27) [ipg_silent, right=0.05cm of p26] {\tiny 1};
        \node (pr27) [ipg_priority, right=0.05cm of pr27] {\tiny 2};
        \node (c27) [ipg_commit, right=0.05cm of pr27] {\tiny 4};
        \node (p26) [xsmall_packet, right=0cm of ipg27] {\tiny 4};
        
        \draw[->] (0.5,1.2) -- (0.5,0.8); 
        \draw[->] (6.125,1.2) -- (6.125,0.8);
        \draw[<->] (0.5 + 0.1,1.0) -- (6.125 - 0.1,1.0);
        \node[] at (0.5, 1.2 + 0.2 + 0.2) {\tiny Packet is};
        \node[] at (0.5, 1.2 + 0.2) {\tiny Ready};
        \node[] at (0.5 + 2.775, 1.0 + 0.2) {\tiny $L_{k=1\mathrm{,}\max}^{\text{PLCA-NP}} = 1.2457$ ms};
        \node[] at (6.125, 1.2 + 0.2 + 0.2) {\tiny Begin};
        \node[] at (6.125, 1.2 + 0.2) {\tiny Transmission};
        
        \end{tikzpicture}
        \label{subfig: Guaranteed maximum access latency for the highest-priority node using PLCA-NP}
    }
    
    \caption{Operation and worst-case access latency of PLCA-NP in an $N=8$ device network, with devices $n=7$ and $n=4$ configured as priority nodes.}
    \label{fig: PLCA-NP operation and access latency in an 8-device network, where devices 7 and 4 configured as priority nodes}

\end{figure}
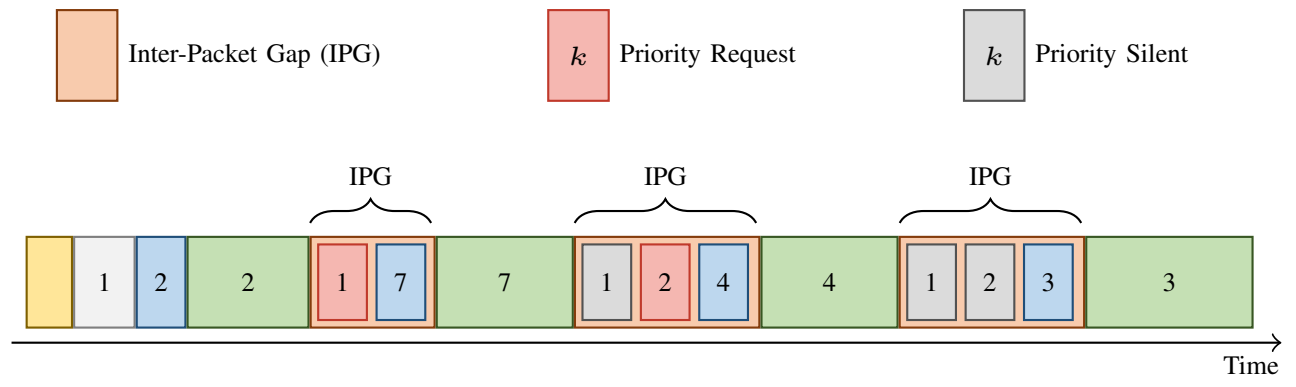
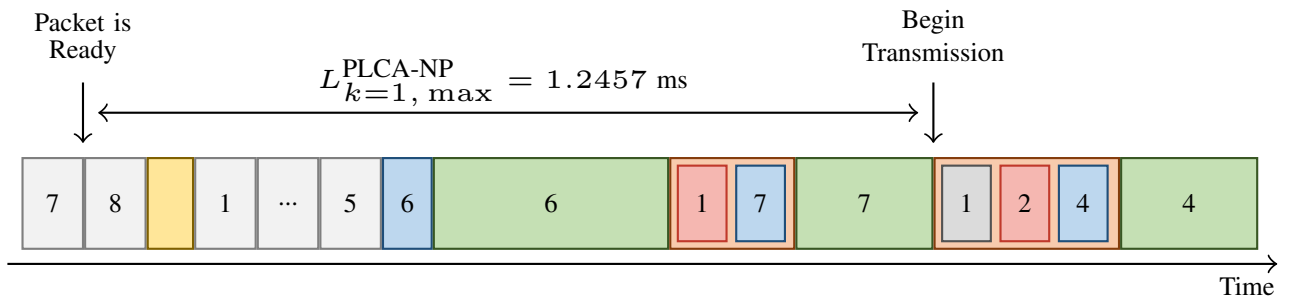

%% file: Ch5_PLCASP.tex
\section{Proposed Priority-Based Network Access and Scheduling Mechanism for PLCA Networks}
\label{Sec: Proposed Mechanism}

This section introduces a priority-based network access mechanism with SP scheduling for PLCA multidrop networks. In contrast to PLCA and PLCA-NP, the proposed mechanism decouples medium access from node-based cyclic ordering and enables class-based access, allowing higher-priority traffic from any device to be transmitted ahead of lower-priority traffic. This provides traffic differentiation with low and bounded access latency while maintaining scalability as the network size increases.

The proposed mechanism is composed of three components, presented successively in this section. Class-based TOs are first introduced to enable priority-based medium access. Building on this, a deterministic ID-based resolution mechanism arranges transmissions among packets of the same priority class. Subsequently, a configuration that emulates SP scheduling is defined. The section concludes with a discussion of implementation considerations as well as the computational and storage requirements.

\subsection{Class-based Transmission Opportunities (CTOs)}
\label{subsec:Priority-Based Network Access}

The proposed mechanism introduces the concept of Class-based Transmission Opportunities (CTOs). A CTO represents a transmission opportunity assigned to a traffic class rather than to a specific device, thereby decoupling medium access from node-based ordering and enabling class-based access.

Each scheduling cycle begins with a BEACON message and consists of $K$ CTOs, one for each traffic class $k \in \{1,\dots, K\}$. The duration of a CTO is adaptive. If no device has pending data for class $k$, CTO-$k$ expires after the silent interval $T_\mathrm{s}$. Otherwise, CTO-$k$ is extended to accommodate the transmission of all data belonging to class $k$.

Devices maintain a local CTO counter, denoted by $\hat{k}$, which indicates the currently active CTO, and use it to determine when to participate based on the priority of their pending packets. When $\hat{k} = k$, devices with pending traffic of class~$k$ attempt to reserve the medium by transmitting a COMMIT message.

Since multiple devices may have pending packets of the same class, simultaneous reservation attempts can occur. If a single device transmits a COMMIT message, the reservation succeeds and the device proceeds with data transmission. If multiple devices transmit COMMIT messages simultaneously, contention\footnote{The proposed mechanism remains collision-free with respect to data transmission, as contention occurs only during channel reservation (COMMIT signaling).} arises within the CTO. In this case, contention is resolved using the deterministic ID-based resolution mechanism described in Section~\ref{subsec: resolution mechanism}.

Once the activity within a CTO is completed---either after a silent interval (when no reservations occur), a single transmission, or the resolution of all contending devices---the CTO counter is incremented and the next CTO begins. Upon completion of the last CTO ($\hat{k} = K$), a new BEACON message is issued to start the next scheduling cycle.

In contrast to PLCA, where the BEACON explicitly marks the end of the node-based cycle, the proposed mechanism does not strictly require a BEACON to signal cycle completion. Since all devices are aware of the number of traffic classes ($K$), the end of a cycle can be inferred locally when $\hat{k} = K$. Nevertheless, the BEACON is retained to ensure robust synchronization in practical deployments, such as device initialization or rejoining during an ongoing cycle. Such unsynchronized devices remain silent until the next BEACON is received, after which they can safely participate in medium access.

\subsection{Deterministic ID-based Resolution}
\label{subsec: resolution mechanism}

Contention may arise when multiple devices simultaneously transmit COMMIT messages to reserve the same CTO. To resolve such contention in a bounded and predictable manner, the proposed mechanism employs a deterministic, ID-based tree resolution mechanism. In contrast to tree-based schemes discussed in Section~\ref{Subsec: Random Access and Collision Resolution}, where branch selection is random, branch selection in the proposed approach is performed deterministically based on device identifier~$n$. This ensures predictable branching behavior and bounded resolution delay.

Let $\mathcal{N}$ denote the set of all devices, with $N = |\mathcal{N}|$, and let $\mathcal{M} \subseteq \mathcal{N}$ denote the set of contending devices, with $M =|\mathcal{M}|$. The number of contending devices is assumed to be known to all devices. Upon detecting a collision, transmissions from non-contending devices are suspended until the collision is resolved. To resolve the collision, the set $\mathcal{N}$ is partitioned into $M$ disjoint subsets (groups), denoted by $\mathcal{S}_1, \mathcal{S}_2, \ldots, \mathcal{S}_{M}$, where $\mathcal{S}_m \subset \mathcal{N}$ and  $m \in \{1, \ldots, M\}$. Partitioning is performed by dividing the ordered set of node IDs into contiguous groups. The groups are constructed to be nearly equal in size: the first $\alpha = N \bmod M$ groups contain $\lceil N / M \rceil$ devices each, while the remaining $(M - \alpha)$ groups contain $\lfloor N / M \rfloor$ devices each.

The groups are granted resolution opportunities sequentially. During the opportunity allocated to group $\mathcal{S}_m$, only devices in $\mathcal{M} \cap \mathcal{S}_m$ attempt to reserve the channel by transmitting COMMIT messages. If no device transmits ($|\mathcal{M} \cap \mathcal{S}_m| = 0$), the opportunity is yielded after the silent interval $T_\mathrm{s}$. If exactly one device transmits ($|\mathcal{M} \cap \mathcal{S}_m| = 1$), the reservation succeeds and the device proceeds with data transmission. If multiple devices transmit ($|\mathcal{M} \cap \mathcal{S}_m| > 1$), a further collision occurs, and the group is recursively subdivided using the same deterministic partitioning procedure. 

After a group is resolved, either by a successful transmission or by remaining silent, the process advances to the next group. The resolution follows a depth-first traversal of the resulting tree and continues until all devices in $\mathcal{M}$ have been resolved. The scheduling cycle then resumes with the next CTO.

\begin{algorithm}[t]
\small
\caption{ID-based Resolution Mechanism}
\label{alg:id_resolution}
\SetAlgoLined
\SetKwFunction{DeterministicSplitting}{DeterministicSplitting}
\SetKwFunction{CollisionResolution}{CollisionResolution}
\SetKwProg{Fn}{Function}{:}{end}
\SetKwInOut{Input}{Input}

\Input{Device set $\mathcal{N}$, contending set $\mathcal{M} \subseteq \mathcal{N}$}

DeterministicSplitting($\mathcal{N}, \mathcal{M}$);

\Fn{\DeterministicSplitting{$\mathcal{N}, \mathcal{M}$}}
{
    $N \gets |\mathcal{N}|$; $M \gets |\mathcal{M}|$\; 
    $\alpha \gets N \bmod M$\;
    $j \gets 1$

    \For{$m \gets 1$ \KwTo $\alpha$}
    {    
        $\mathcal{S}_m \gets \{n_j, \ldots, n_{j+\lceil N/M \rceil - 1}\}$\;
        $j \gets j + \lceil N/M \rceil$\;
    }
    \For{$m \gets \alpha+1$ \KwTo $M$}
    {
        $\mathcal{S}_m \gets \{n_j, \ldots, n_{j+\lfloor N/M \rfloor -1}\}$\;
        $j \gets j + \lfloor N/M \rfloor$\;
    }
    CollisionResolution($\{\mathcal{S}_1,\ldots,\mathcal{S}_M\}, \mathcal{M}$)\;
}
\Fn{\CollisionResolution{$\{\mathcal{S}_1,\ldots,\mathcal{S}_M\}, \mathcal{M}$}}
{
    \For{$m \gets 1$ \KwTo $M$}
    {

        \If{$|\mathcal{M} \cap \mathcal{S}_m| = 0$}
        {
            Idle interval of duration $T_\mathrm{s}$\;
        }

        \ElseIf{$|\mathcal{M} \cap \mathcal{S}_m| = 1$}
        {
            Successful transmission by the device in $\mathcal{M} \cap \mathcal{S}_m$\;
        }

        \Else
        {
            Collision interval of duration $T_\mathrm{c}$\;
            DeterministicSplitting($\mathcal{S}_m, \mathcal{M} \cap \mathcal{S}_m$)\;
        }
    }

}

\end{algorithm}

The operation of the ID-based resolution mechanism is presented in Algorithm~\ref{alg:id_resolution} and further illustrated in Fig.~\ref{fig: Transmission behavior under the proposed priority-based network access with ID-based resolution in an 8-device network, where devices 1, 3, and 8 transmit during CTO-1 and devices 4 and 5 transmit during CTO-2} for an $N=8$ device network. The figure shows both the transmission behavior within a CTO (Fig. \ref{fig: Transmission behavior of the proposed priority-based access mechanism}) and the corresponding resolution processes (Figs. \ref{fig: Collision resolution during CTO-1 using the ID-based tree mechanism} and \ref{fig: Collision resolution during CTO-2 using the ID-based tree mechanism}).

In this example, devices $n=1$, $3$, and $8$ contend for CTO-1, while devices $n=4$ and $5$ contend for CTO-2. In CTO-1 (Fig. \ref{fig: Collision resolution during CTO-1 using the ID-based tree mechanism}), the initial collision occurs over the full ID space $\mathcal{N} = \{1,2,\ldots,8\}$ with contending set $\mathcal{M} = \{1,3,8\}$. Since $M = 3$, the set $\mathcal{N}$ is partitioned into three groups: $\mathcal{S}_1 = \{1,2,3\}$, $\mathcal{S}_2 = \{4,5,6\}$, and $\mathcal{S}_3 = \{7,8\}$.

During the first resolution opportunity, only devices in $\mathcal{M} \cap \mathcal{S}_1 = \{1,3\}$ attempt transmission, resulting in a further collision. Group $\mathcal{S}_1$ is then subdivided into $\mathcal{S}_{1.1} = \{1,2\}$ and $\mathcal{S}_{1.2} = \{3\}$. In the subsequent opportunity, allocated to $\mathcal{S}_{1.1}$, device $n=1$ successfully reserves the channel and transmits its data, followed by device $n=3$ in the opportunity allocated to $\mathcal{S}_{1.2}$. 

The process then proceeds to $\mathcal{S}_2$, where no devices contend ($\mathcal{M} \cap \mathcal{S}_2 = \emptyset$), resulting in a silent interval. Finally, device $n=8$ transmits during the opportunity allocated to $\mathcal{S}_3$, completing the resolution and allowing the cycle to advance to CTO-2.

\input{Fig3_PNA}

\subsection{Strict Priority (SP) Scheduling}
\label{sec:Emulation of Strict Priority Scheduling}

The proposed mechanism can be configured to emulate SP scheduling. This configuration, referred to as PLCA-SP, enforces strict priority ordering across traffic classes by resetting the CTO counter to the highest-priority class ($\hat{k} = 1$) before completing the full scheduling cycle.

After any successful data transmission in a lower-priority CTO ($\hat{k} > 1$), the CTO counter is immediately reset to ($\hat{k} = 1$). For the highest-priority CTO ($\hat{k} = 1$), the reset occurs only after the corresponding resolution process has been fully resolved, i.e., after all contending devices have successfully transmitted. As a result, a CTO of class $k$ is served only if all higher-priority classes have no pending traffic.

The SP configuration contrasts with the default operation described in Section~\ref{subsec:Priority-Based Network Access}. In the default configuration, the CTO counter is incremented only after contention within the current CTO is fully resolved and is reset only after the final CTO ($\hat{k} = K$), resulting in a cyclic service order across priority classes rather than strict priority scheduling. However, BEACON transmission remains unchanged: a BEACON is transmitted only upon completion of the final CTO ($\hat{k}=K$). Additional resets of the CTO counter to $\hat{k}=1$ under SP scheduling do not trigger BEACON transmission.

A comparison between the default and SP configurations is shown in Fig.~\ref{fig: Comparison between default and SP configuration}. In the considered example (Fig.~\ref{subfig: The considered Example}), packets of class $k=2$ are pending at devices $n=3$ and $7$, and a packet of class $k=3$ is pending at device $n=4$. A packet of class $k=1$ arrives at device $n=2$ after the first occurrence of CTO-1 has elapsed. This packet, under the default configuration (Fig.~\ref{subfig: Transmission behavior under the default configuration}), must wait until the next scheduling cycle before gaining access, whereas under the SP configuration (Fig.~\ref{subfig: Transmission behavior under the SP configuration}), it gains access immediately after the current transmission completes.

Since, under the SP configuration, any successful transmission in a lower-priority CTO resets the CTO counter, resolution processes may span multiple scheduling cycles. A critical design aspect is therefore how such interrupted resolution processes are handled. We consider two approaches:

\input{Fig4_Comparison_of_default_and_SP_configuration}

\subsubsection{Stop-and-Restart (SR)}
In the Stop-and-Restart approach, resolution proceeds until the first successful transmission, after which it is terminated and the CTO counter is reset to $\hat{k} = 1$. When the corresponding CTO is revisited, a new resolution process is initiated. Devices that were previously contending compete again together with newly arriving devices, without preserving their prior position in the resolution process. This trades fairness for simplicity, as no contention state needs to be maintained across cycles.

\subsubsection{Pause-and-Resume (PR)}
In the Pause-and-Resume approach, resolution proceeds until the first successful transmission, after which it is paused rather than terminated, and resumed when the corresponding CTO is revisited. Newly arriving packets wait until the current resolution tree is fully resolved before joining a subsequent resolution process. This preserves contention order across cycles and enables First-Come-First-Served (FCFS) service discipline between successive resolution batches within the same class. While PR improves fairness among devices within the same class, it requires additional overhead to maintain the resolution state.

Crucially, both SR and PR yield identical behavior for the highest-priority class. In both cases, any successful transmission in a lower-priority CTO resets the CTO counter to $\hat{k}=1$, ensuring the same access pattern and worst-case latency guarantees for the highest-priority traffic. Their impact is limited to contention management within lower-priority CTOs.

\subsection{Implementation Considerations and Complexity}
\label{sec:Implementation Considerations and Complexity}

The proposed mechanism requires extensions to the PLCA Reconciliation Sublayer behavior defined in IEEE 802.3cg, Clause 148. In standard PLCA, the PLCA Data function signals that the MAC has a packet pending for transmission by asserting the \textit{packetPending} flag. The PLCA Control function uses this indication to commit to the TO assigned to the device. In the proposed mechanism, TOs are associated with traffic classes. The PLCA Control function must therefore determine whether eligible traffic is available for the current CTO. This requires coordination between packet selection and medium access to ensure that the selected packet belongs to the class being served.

Class-based access allows  multiple devices with pending traffic of the same class to attempt channel reservation simultaneously, potentially resulting in contention. Resolving this contention through the proposed ID-based tree mechanism requires distinguishing idle, successful, and colliding reservation outcomes and determining the number of contenders involved in each collision. This work assumes that these outcomes are identified without error and that the exact contender count is available to all devices.

The computational and storage requirements of the ID-based tree resolution mechanism can be characterized using a compact implementation in which contiguous groups are represented by their boundaries and subgroup boundaries are generated as needed. Assuming constant-time arithmetic on identifiers and counters, completing one resolution tree requires $O(N)$ operations per device. The leaf groups partition the ID space into at most $N$ nonempty groups, and each internal node has at least two children, corresponding to subgroups, yielding at most $2N-1$ tree nodes. For storage, a depth-first traversal retains a constant number of machine words per active level. Each subdivision of a colliding group $\mathcal{S}_m$ produces subgroups of size at most $\lceil |\mathcal{S}_m|/2\rceil$. The tree depth is therefore at most $\lceil\log_2 N\rceil$, giving $O(\log_2 N)$ machine words of traversal storage per device. Under PR, maintaining separate resolution states for up to $K$ classes requires $O(K\log_2 N)$ machine words per device.

%% file: Fig3_PNA.tex
\input{DefineStyles_P3}
\begin{figure}[!t]
    \centering
    \subfloat[Resolution process for contending devices $n=1,3,8$ during CTO-1.]{ 
        \begin{tikzpicture}       
        [
            level distance=1.5cm, 
            sibling distance=2cm 
        ]

        \node[draw=PriorityOL, fill=PriorityFill, circle, minimum size=0.8cm] {\tiny $1,3,8$}
            child [draw=WhiteOLL]{node[draw=PriorityOL, fill=PriorityFill, circle, minimum size=0.8cm] {\tiny $1,3$} 
                child[draw=WhiteOLL] {node[draw=BlueOLD, fill=BlueFillD, circle, minimum size=0.8cm] {\tiny $1$}} 
                child[draw=WhiteOLL] {node[draw=BlueOLD, fill=BlueFillD, circle, minimum size=0.8cm] {\tiny $3$}} 
            }
            child [draw=WhiteOLL]{node[draw=WhiteOLD, fill=WhiteFillL, circle, minimum size=0.8cm] {}}
            child[draw=WhiteOLL] {node[draw=BlueOLD, fill=BlueFillD, circle, minimum size=0.8cm] {\tiny $8$}};

            \node[] at (0,0 + 0.6 + 0.25) {\tiny Set $\mathcal{N} =$};
            \node[] at (0,0 + 0.6) {\tiny $\{1,2,\dots,8\}$};

            \node[] at (0 - 2,-1.5 + 0.5 + 0.25) {\tiny Set $1 =$};
            \node[] at (0 - 2,-1.5 + 0.5) {\tiny $\{1,2,3\}$};

            \node[] at (0,-1.5 + 0.5 + 0.25) {\tiny Set $2 =$};
            \node[] at (0,-1.5 + 0.5) {\tiny $\{4,5,6\}$};

            \node[] at (0 + 2,-1.5 + 0.5 + 0.25) {\tiny Set $3 =$};
            \node[] at (0 + 2,-1.5 + 0.5) {\tiny $\{7,8\}$};

            \node[] at (0 - 2 - 1,0 -1.5 -1.5 + 0.5 + 0.25) {\tiny Set $1.1 =$};            \node[] at (0 - 2 - 1,0 -1.5 -1.5 + 0.5) {\tiny $\{1,2\}$};

            \node[] at (0 - 2 + 1,0 -1.5 -1.5 + 0.5 + 0.25) {\tiny Set $1.2 =$};        
            \node[] at (0 - 2 + 1,0 -1.5 -1.5 + 0.5) {\tiny $\{3\}$};

            \node [] at (0 - 2 - 1 -1,0 -1.5 -1.5 - 0.25) {\color{white} \tiny .};
            \node [] at (0 + 2 + 1 + 1,0 -1.5 -1.5 - 0.25) {\color{white} \tiny .};

        \end{tikzpicture}
        \label{fig: Collision resolution during CTO-1 using the ID-based tree mechanism}
 
    }

    \subfloat[Resolution process for contending devices $n=4,5$ during CTO-2.]{ 
    \begin{tikzpicture}       
    [
        level distance=1.5cm, 
        sibling distance=2cm 
    ]

    \node[draw=PriorityOL, fill=PriorityFill, circle, minimum size=0.8cm] {\tiny $4,5$}
        child [draw=WhiteOLL]{node[draw=BlueOLD, fill=BlueFillD, circle, minimum size=0.8cm] {\tiny $4$}}
        child[draw=WhiteOLL] {node[draw=BlueOLD, fill=BlueFillD, circle, minimum size=0.8cm] {\tiny $5$}};

        \node[] at (0,0 + 0.6 + 0.25) {\tiny Set $\mathcal{N} =$};
        \node[] at (0,0 + 0.6) {\tiny $\{1,2,\dots,8\}$};

        \node[] at (0 - 1,-1.5 + 0.5 + 0.25) {\tiny Set $1 =$};
        \node[] at (0 - 1,-1.5 + 0.5) {\tiny $\{1,\dots,4\}$};

        \node[] at (0 + 1,-1.5 + 0.5 + 0.25) {\tiny Set $2 =$};
        \node[] at (0 + 1,-1.5 + 0.5) {\tiny $\{5,\dots,8\}$};

        \node [] at (0 - 2 - 1 -1,0 -1.5 - 0.25) {\color{white} \tiny .};
        \node [] at (0 + 2 + 1 + 1,0 -1.5 - 0.25) {\color{white} \tiny .};

    \end{tikzpicture}
    \label{fig: Collision resolution during CTO-2 using the ID-based tree mechanism}

    }

   \subfloat[CTO-based transmission sequence.]{%
        \begin{tikzpicture}
        \draw[->] (0,0.8) -- (8.4,0.8) node[right]{};

        \node at (8.2, 0.8+0.15) {\tiny Time};

        
        \node (b20) [beacon] at (0.25,0.4) {};
        \node(p10) [priority, right=0cm of b20] {\tiny 1};
        \node(c11) [priority, right=0cm of p10] {\tiny 1};
        \node(c12) [commit, right=0cm of c11] {\tiny 1};
        \node (p13) [min_packet, right=0cm of c12] {\tiny 1};
        
        \node(c21) [commit, right=0cm of p13] {\tiny 3};
        \node (p22) [min_packet, right=0cm of c21] {\tiny 3};

        \node (s31) [silent, right=0cm of p22] {\tiny 1};

        \node(c41) [commit, right=0cm of s31] {\tiny 8};
        \node (p42) [min_packet, right=0cm of c41] {\tiny 8};

        \node(p50) [priority, right=0cm of p42] {\tiny 2};
        \node(c51) [commit, right=0cm of p50] {\tiny 4};
        \node (p52) [min_packet, right=0cm of c51] {\tiny 4};

        \node(c61) [commit, right=0cm of p52] {\tiny 5};
        \node (p62) [min_packet, right=0cm of c61] {\tiny 5};

        \node[above=-0.2cm of p10] (card17) [CTO_commit, xshift=+0.2cm] {\tiny 8};
        \node[above=0.05cm of p10] (card13) [CTO_commit] {\tiny 3};
        \node[above=-0.2cm of p10] (card12) [CTO_commit, xshift=-0.2cm] {\tiny 1};

        \node[above=-0.1cm of c11] (card22) [CTO_commit] {\tiny 1};
        \node[above=0.1cm of c11] (card22) [CTO_commit, xshift=0.2cm] {\tiny 3};

        \node[above=-0.1cm of p50] (card34) [CTO_commit] {\tiny 4};
        \node[above=0.1cm of p50] (card35) [CTO_commit, xshift=0.2cm] {\tiny 5};

        \draw[decorate, decoration={brace, amplitude=1ex, raise=1ex}] (5.5,0.15)  -- ++ (-5.075,0);
        \node [] at (5.5 - 2.5875, 0.15-0.5) {\scriptsize CTO-1};

        \draw[decorate, decoration={brace, amplitude=1ex, raise=1ex}] (8.385,0.15)  -- ++ (-2.9,0);
        \node [] at (8.385 - 1.45, 0.15-0.5) {\scriptsize CTO-2};

        \begin{scope}[shift={(0, 1.9)}] 
        
        \node () [beaconStd] at (0.25+0.25,0) {};
        \node[right] at (0.25+0.15+0.25, 0) {\tiny BEACON};

        \node () [commitStd] at (2+0.25,0) {\tiny $n$};
        \node[right] at (2+0.15+0.25, 0) {\tiny COMMIT};
        
        \node () [ipg_priorityStd] at (3.75+0.25,0) {\tiny $k$};
        \node[right] at (3.75+0.15+0.25, 0) {\tiny Class};

        \node () [silentStd] at (5.375+0.25,0) {\tiny $k$};
        \node[right] at (5.375+0.15+0.25, 0) {\tiny Silent};

        \node () [dataStd] at (7+0.25,0) {\tiny $n$};
        \node[right] at (7+0.15+0.25, 0) {\tiny Data};
        \end{scope}

        \end{tikzpicture}
        \label{fig: Transmission behavior of the proposed priority-based access mechanism}
    }
    
    \caption{Operation of the proposed priority-based access mechanism in an $N=8$ device network, illustrating CTO-based transmission and the corresponding resolution processes. Devices $n=1, 3, 8$ contend during CTO-1, while devices $n=4,5$ contend during CTO-2.}
    \label{fig: Transmission behavior under the proposed priority-based network access with ID-based resolution in an 8-device network, where devices 1, 3, and 8 transmit during CTO-1 and devices 4 and 5 transmit during CTO-2}

\end{figure}
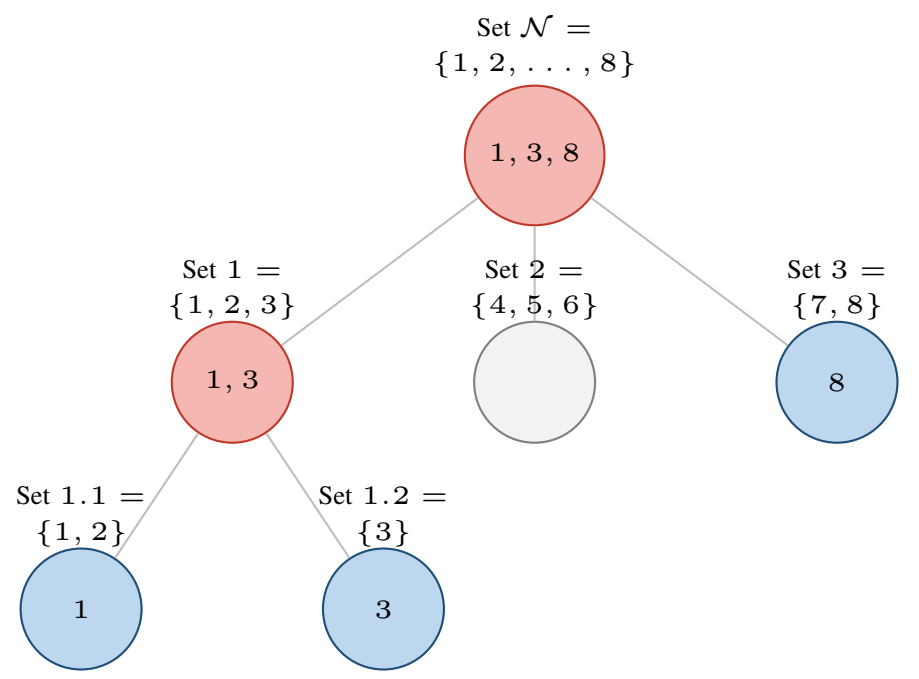
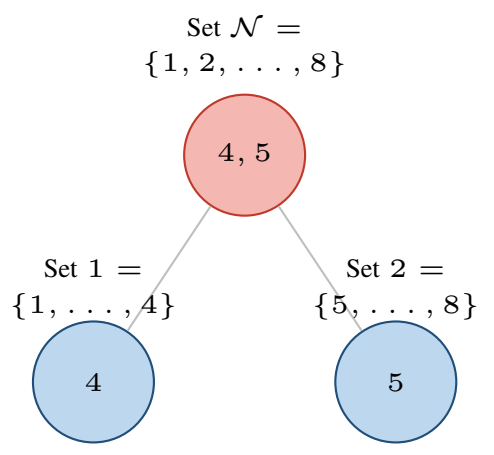
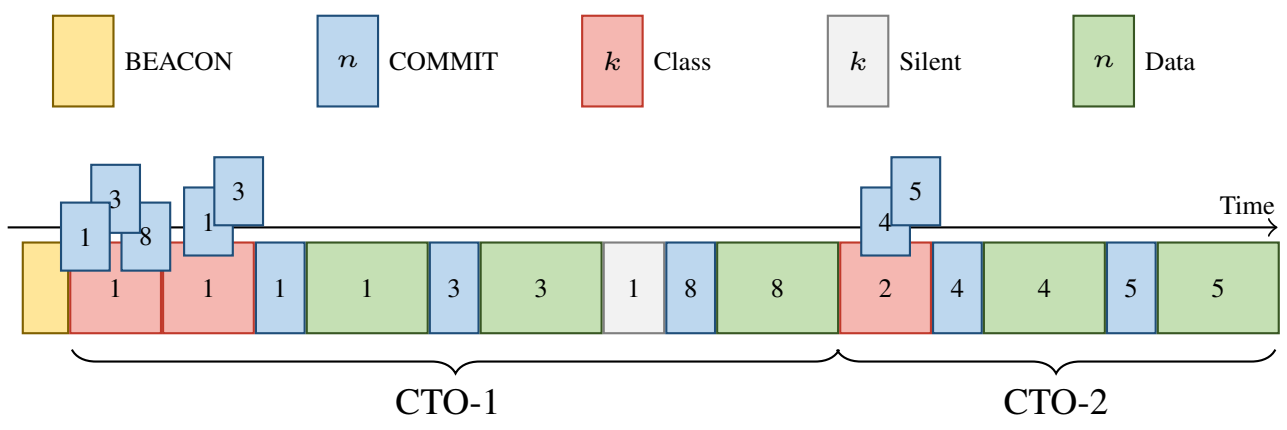

%% file: Fig4_Comparison_of_default_and_SP_configuration.tex
\def\h{0.9}
\newcommand{\drawQueue}[3]{
    \foreach \i in #1 {
        \foreach \j in #2 {
        \draw[draw=gray, thick] (\i*4 - 3.95, \j*\h) rectangle (\i*4 - 3.9 + #3*0.39, \j*\h + 0.4);
        \pgfmathsetmacro{\w}{#3/2}
        }
    }
}

\begin{figure}
    \centering
    \subfloat[Considered traffic scenario: class-$2$ packets at devices $n=3,7$, a class-$3$ packet at device $n=4$, and a class-$1$ packet arriving at device $n=2$ after the first occurrence of CTO-1.]{
        \begin{tikzpicture}[]

        \draw[] (1,0) -- (7,0);

        \foreach \x/\i in {0/1, 2/3,  5/7} {

        \draw[] (\x+1,0) -- (\x+1,0.5);

        \draw[rounded corners] (\x+0.5,0.5) rectangle (\x+1.5,0.5+0.6);

        \node at (\x+1,0.5+0.3) {\scriptsize $n = $~\i};

        \draw[draw=gray, thick] (\x+0.25,1.75) -- (\x+1.75,1.75); 
        \draw[draw=gray, thick] (\x+1.75,1.75) -- (\x+1.75,1.25); 
        \draw[draw=gray, thick] (\x+1.75,1.25) -- (\x+0.25,1.25); 
        }

        \node at (3.5+1,0.5+0.3) {\scriptsize $\ldots$};

        \foreach \x/\i in {1/2, 3/4, 6/8} {
        
        \draw[] (\x+1,0) -- (\x+1,-0.5);
        
        \draw[rounded corners] (\x+0.5,-0.5) rectangle (\x+1.5,-0.5-0.6);
        
        \node at (\x+1,-0.5-0.3) {\scriptsize $n = $~\i};

        \draw[draw=gray, thick] (\x+0.25,-1.75) -- (\x+1.75,-1.75); 
        \draw[draw=gray, thick] (\x+1.75,-1.75) -- (\x+1.75,-1.25); 
        \draw[draw=gray, thick] (\x+1.75,-1.25) -- (\x+0.25,-1.25); 
        }

        \node() [pkt] at (2+1.725-0.4,1.5) {\tiny $k=2$};
        \node() [pkt] at (5+1.725-0.4,1.5) {\tiny $k=2$};
        \node() [pkt] at (3+1.725-0.4,-1.5) {\tiny $k=3$};
        \node() [pkt] at (1-0.4,-1.5) {\tiny $k=1$};
        \draw[->] (1+0.04,-1.5) -- (1+0.44,-1.5);

        \node at (4.5+1,-0.5-0.3) {\scriptsize $\ldots$};
        \label{subfig: The considered Example}
        
        \end{tikzpicture}
    }

   \subfloat[Transmission sequence under the default configuration.]{%
        \begin{tikzpicture}

        \draw[->] (0,0.8) -- (8.4,0.8) node[right]{};

        \node at (8.2, 0.8+0.15) {\tiny Time};

        \node (b20) [beacon] at (0.25,0.4) {};
        \node(p10) [priority2, right=0cm of b20] {\tiny 1};

        \node[above=-0.1cm of p10] () [CTO_silent] {\tiny 1};
        
        \node(p20) [priority2, right=0cm of p10] {\tiny 2};
        \node(c21) [commit, right=0cm of p20] {\tiny 3};
        \node (p21) [small_packet, right=0cm of c21] {\tiny 3};

        \node[above=-0.1cm of p20] [CTO_commit] {\tiny 3};
        \node[above=0.1cm of p20] () [CTO_commit, xshift=0.2cm] {\tiny 7};
        
        \node(c22) [commit, right=0cm of p21] {\tiny 7};
        \node (p22) [large_packet, right=0cm of c22] {\tiny 7};

        \node(p30) [priority2, right=0cm of p22] {\tiny 3};

        \node (p32) [large_packet, right=0cm of p30] {\tiny 4};

        \node[above=-0.1cm of p30] () [CTO_commit] {\tiny 4};

        \node (b40) [beacon, right=0cm of p32]{};

        \node(p40) [priority2, right=0cm of b40] {\tiny 1};
        \node[above=-0.1cm of p40] (card34) [CTO_commit] {\tiny 2};

        \node (p62) [small_packet, right=0cm of p40] {\tiny 2};

        \draw[->] (0.825,-0.4) -- (0.825,0); 
        \draw[->] (6.5,-0.4) -- (6.5,0);
        \draw[<->] (0.825 + 0.1,-0.2) -- (6.5 - 0.1,-0.2);
        \node[] at (0.825 + 0.01, -0.4 - 0.2) {\tiny Packet};
        \node[] at (0.825 + 0.02, -0.4 - 0.2 - 0.2) {\tiny is Ready};

        \node[] at (6.5, -0.4 - 0.2) {\tiny Begin};
        \node[] at (6.5, -0.4 - 0.2 - 0.2) {\tiny Transmission};

        \begin{scope}[shift={(0, 1.9)}] 
        
        \node () [beaconStd] at (0.25+0.25,0) {};
        \node[right] at (0.25+0.15+0.25, 0) {\tiny BEACON};

        \node () [commitStd] at (2+0.25,0) {\tiny $n$};
        \node[right] at (2+0.15+0.25, 0) {\tiny COMMIT};
        
        \node () [ipg_priorityStd] at (3.75+0.25,0) {\tiny $k$};
        \node[right] at (3.75+0.15+0.25, 0) {\tiny Class};

        \node () [silentStd] at (5.375+0.25,0) {\tiny $k$};
        \node[right] at (5.375+0.15+0.25, 0) {\tiny Silent};

        \node () [dataStd] at (7+0.25,0) {\tiny $n$};
        \node[right] at (7+0.15+0.25, 0) {\tiny Data};
        \end{scope}

        \end{tikzpicture}
        \label{subfig: Transmission behavior under the default configuration}
    }

   \subfloat[Transmission sequence under the SP configuration (PLCA-SP).]{%
        \begin{tikzpicture}
        
        \draw[->] (0,0.8) -- (8.4,0.8) node[right]{};

        \node at (8.2, 0.8+0.15) {\tiny Time};

        \node (b20) [beacon] at (0.25,0.4) {};
        
        \node(p10) [priority2, right=0cm of b20] {\tiny 1};
        \node[above=-0.1cm of p10] () [CTO_silent] {\tiny 1};
        
        \node(p20) [priority2, right=0cm of p10] {\tiny 2};
        \node(c21) [commit, right=0cm of p20] {\tiny 3};
        \node (p21) [small_packet, right=0cm of c21] {\tiny 3};

        \node[above=-0.1cm of p20] () [CTO_commit] {\tiny 3};
        \node[above=0.1cm of p20] () [CTO_commit, xshift=0.2cm] {\tiny 7};

        \node(p40) [priority2, right=0cm of p21] {\tiny 1};
        \node[above=-0.1cm of p40] (card34) [CTO_commit] {\tiny 2};
        \node (p62) [small_packet, right=0cm of p40] {\tiny 2};

        \node(p50) [priority2, right=0cm of p62] {\tiny 1};
        \node[above=-0.1cm of p50] () [CTO_silent] {\tiny 1};

        \node(p60) [priority2, right=0cm of p50] {\tiny 2};
        \node[above=-0.1cm of p60] (card22) [CTO_commit] {\tiny 7};
        \node (p22) [ave_packet, right=0cm of p60] {\tiny 7};

        \node(p100) [priority2, right=0cm of p22] {\tiny 1};
        \node[above=-0.1cm of p100] () [CTO_silent] {\tiny 1};
        \node(p200) [priority2, right=0cm of p100] {\tiny 2};
        \node[above=-0.1cm of p200] () [CTO_silent] {\tiny 2};

        \node(p30) [priority2, right=0cm of p200] {\tiny 3};
        
        \node (p32) [ave_packet, right=0cm of p30] {\tiny 4};

        \node[above=-0.1cm of p30] (card34) [CTO_commit] {\tiny 4};

        \draw[->] (0.825,-0.4) -- (0.825,0); 
        \draw[->] (3,-0.4) -- (3,0);
        \draw[<->] (0.85 + 0.1,-0.2) -- (3 - 0.1,-0.2);
        \node[] at (0.825 + 0.01, -0.4 - 0.2) {\tiny Packet};
        \node[] at (0.825 + 0.02, -0.4 - 0.2 - 0.2) {\tiny is Ready};
 
        \node[] at (3, -0.4 - 0.2) {\tiny Begin};
        \node[] at (3, -0.4 - 0.2 - 0.2) {\tiny Transmission};

        \end{tikzpicture}
        \label{subfig: Transmission behavior under the SP configuration}
    }
        \caption{Comparison of transmission behavior under the default configuration and Strict-Priority configuration (PLCA-SP) for the proposed mechanism.}
        \label{fig: Comparison between default and SP configuration}

\end{figure}
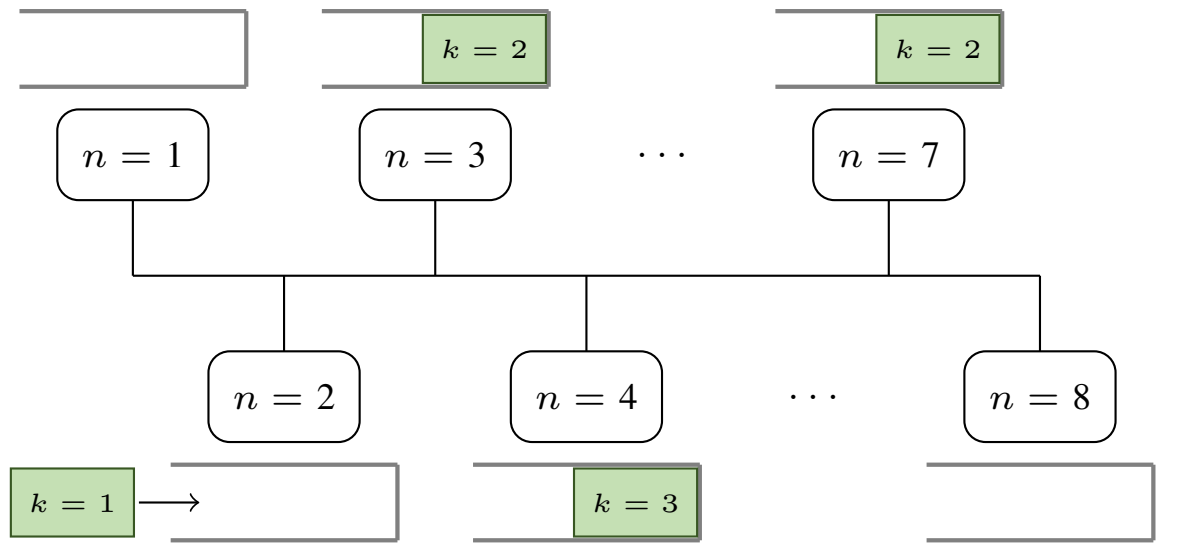
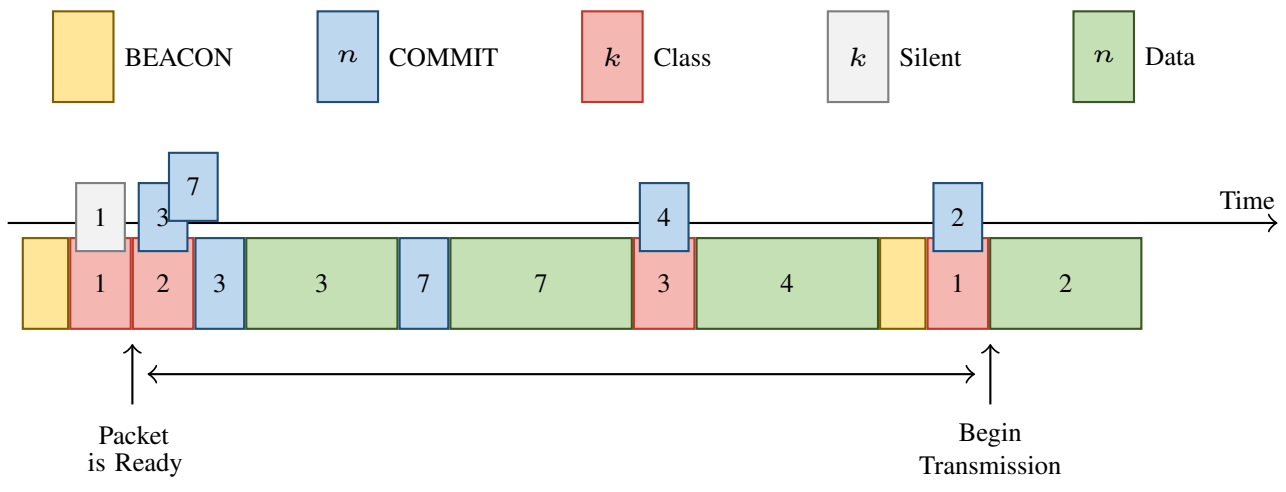
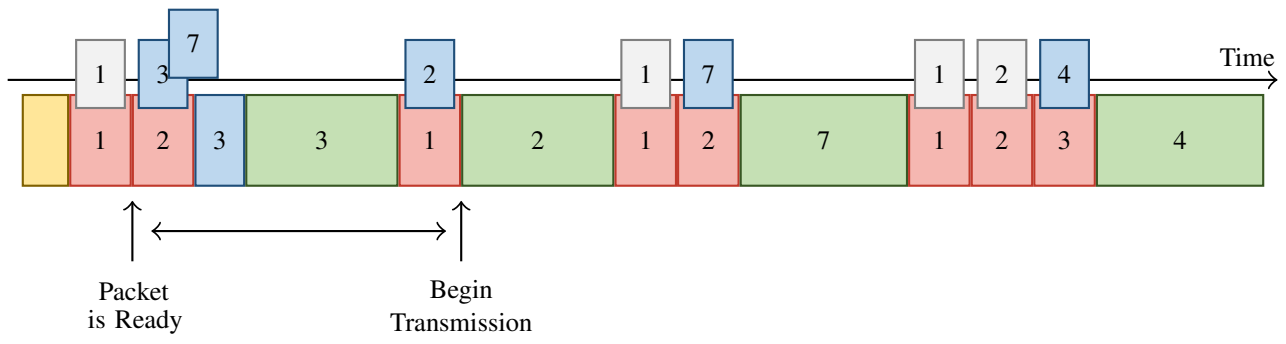

%% file: Ch6_derivations.tex
\section{Worst-Case Latency Analysis of the Proposed Mechanism under the SP Configuration}
\label{sec:Worst-Case Access Latency Analysis under SP Configuration}

This section analyzes the worst-case latency of the highest-priority traffic ($k = 1$) under the SP configuration introduced in Section~\ref{Sec: Proposed Mechanism}. Without additional constraints on higher-priority traffic, lower-priority classes have no finite worst-case latency guarantee, since continuously pending higher-priority traffic can indefinitely defer their service. The latency is primarily determined by the behavior of the resolution mechanism. Accordingly, the analysis proceeds in two steps. First, we characterize the latency induced by the resolution process. Second, we derive the resulting worst-case latency bounds.

\subsection{Latency Induced by the Resolution Process}
\label{subsubsec:treelatency}

We begin by characterizing the latency induced by the resolution mechanism. As specified in Section~\ref{subsec:system model}, the $N$ devices have unique, contiguous node identifiers from $1$ to $N$. Thus, the address space of the resolution mechanism has size $N$, equal to the number of devices. The resolution process can be represented as a traversal of a tree structure. In this representation, collisions form internal nodes, indicating that multiple devices remain contending ($|\mathcal{M} \cap \mathcal{S}_m| > 1$) and that further subdivision is required. Silent intervals form empty leaf nodes, where no contending devices are present ($|\mathcal{M} \cap \mathcal{S}_m| = 0$), while successful transmissions form singleton leaf nodes corresponding to exactly one contending device ($|\mathcal{M} \cap \mathcal{S}_m| = 1$).

The latency introduced by the resolution mechanism depends on the extent to which the resolution tree is traversed. We consider three traversal scenarios: (1) \textit{Full Tree (FT)}, in which the entire resolution tree is explored; (2) \textit{Last Success (LS)}, where traversal stops after the last successful transmission; and (3) \textit{First Success (FS)}, where traversal terminates upon the first successful transmission. The distinction between FT and LS is that FT includes silent intervals occurring after the final successful transmission, whereas LS excludes them.

For each scenario $x \in \{\mathrm{FT}, \mathrm{LS}, \mathrm{FS}\}$, we derive four quantities: 
\begin{itemize}
    \item the maximum number of idle intervals $I_{\max}^{x}$;
    \item the maximum number of collisions $C_{\max}^{x}$;
    \item the maximum number of idle intervals under maximum collisions $I_{C,\max}^{x}$;
    \item the maximum number of collisions under maximum idle intervals $C_{I,\max}^{x}$.
\end{itemize}
Idle intervals (i.e., $I_{\max}^{x}$ and $I_{C,\max}^{x}$) correspond to empty leaf nodes in the resolution tree, whereas collisions (i.e., $C_{\max}^{x}$ and $C_{I,\max}^{x}$) correspond to internal nodes.

\begin{figure*}[!t]  
    \centering 
    \includegraphics[width=\linewidth]{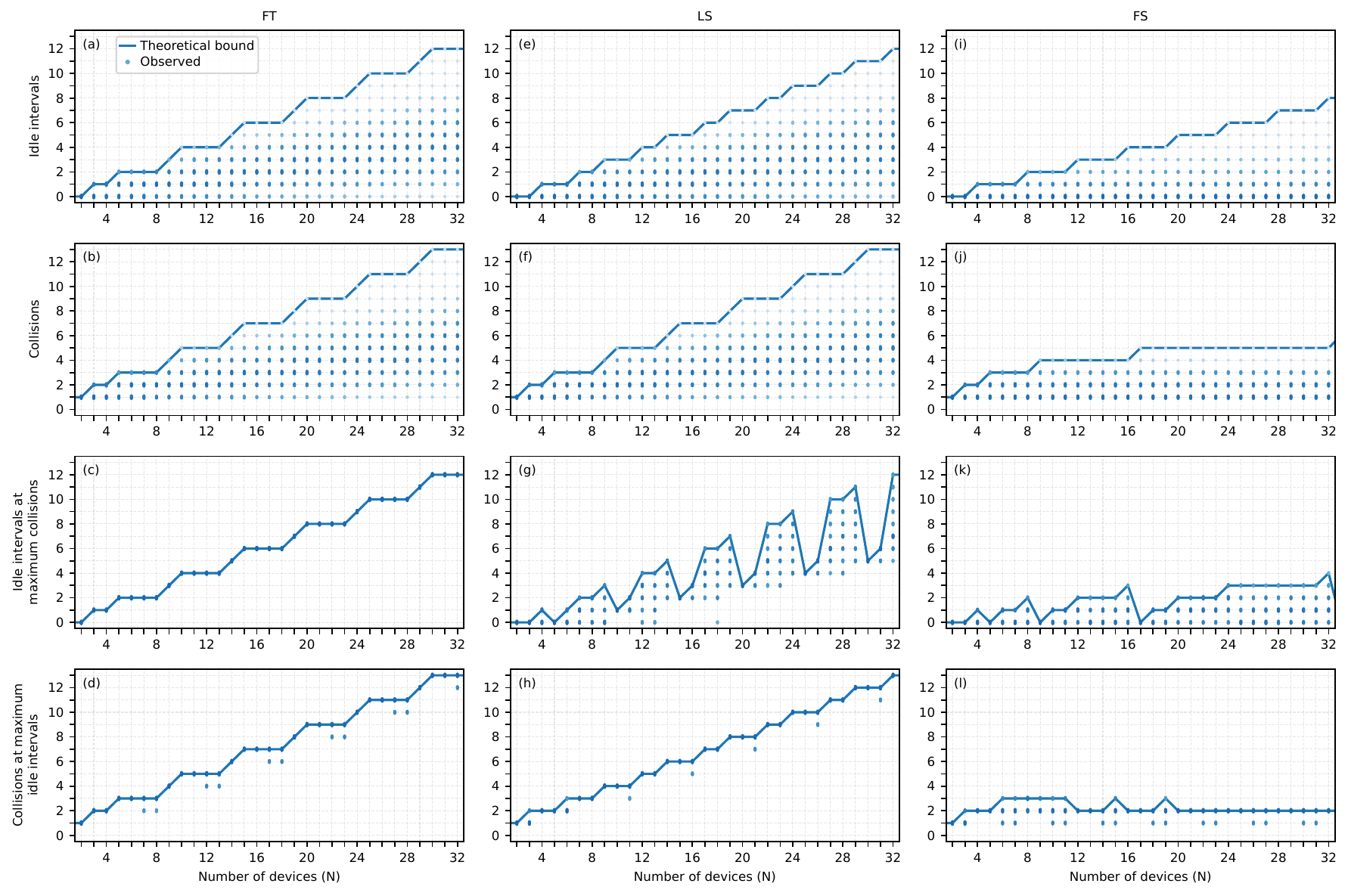}
    \caption{Bounds on the numbers of idle intervals and collisions, and corresponding observed values obtained through exhaustive enumeration for the Full Tree (FT), Last Success (LS), and First Success (FS) traversal scenarios, for network sizes $N=2$--$32$.}
    \label{fig: validation of analytical tree bounds} 
\end{figure*}

These bounds form the foundation for the worst-case latency analysis presented in Section \ref{subsubsection: latency bounds}. Their validation is provided in Section~\ref{subsec: Validation of Resolution Latency Bounds}, and their behavior can be visualized in Fig.~\ref{fig: validation of analytical tree bounds}.

\subsubsection{Full Tree (FT) Scenario}
The FT scenario characterizes the resolution process under complete traversal of the contention tree, including silent intervals occurring after the final successful transmission. Under the FT scenario, the maximum number of idle intervals, $I_{\max}^{\mathrm{FT}}$, coincides with the maximum number of idle intervals under maximum collisions, $I_{C,\max}^{\mathrm{FT}}$, and is expressed as
\begin{equation}
I_{\max}^{\mathrm{FT}} = I_{C\mathrm{,}\max}^{\mathrm{FT}} =
\begin{cases}
0, & N = 2, \\
1, & N = 3, \\
2  \left\lfloor \frac{N}{5} \right\rfloor + 1_{\{p=4\}} & N \geq 4,
\end{cases}
\label{equ:I_max of FT}
\end{equation}
where $p = N \bmod 5$.

For $N = 2$, contention is resolved immediately and no idle intervals occur, whereas for $N = 3$, a single idle interval separates the colliding devices. More generally, for networks with $N \geq 4$, $I_{\max}^{\mathrm{FT}}$ and $I_{C\mathrm{,}\max}^{\mathrm{FT}}$ exhibit a step-wise growth with increasing network size, as shown in Figs.~\ref{fig: validation of analytical tree bounds}a and~\ref{fig: validation of analytical tree bounds}c. The term $2\lfloor N/5 \rfloor$ captures the periodic increase in idle intervals, while the indicator term $1_{\{p=4\}}$ accounts for the intermediate step preceding the completion of each period.

The worst-case behavior occurs when collisions involve pairs of adjacent devices in the ID space, with each colliding pair assigned to a distinct group at the first level of the resolution tree, particularly among the first  $\lfloor M/2 \rfloor$ groups, $\mathcal{S}_{1},\dots,\mathcal{S}_{\lfloor M/2 \rfloor} $. Due to the partitioning mechanism described in Section \ref{subsec: resolution mechanism}, the first $\alpha$ groups may contain more devices than the remaining groups. Assigning colliding pairs to these larger groups maximizes the accumulation of idle intervals throughout the resolution process.

Similarly, under the FT scenario, the maximum number of collisions, $C_{\max}^{\mathrm{FT}}$, coincides with the maximum number of collisions under maximum idle intervals, $C_{I,\max}^{\mathrm{FT}}$, and satisfies
\begin{equation}
C_{\max}^{\mathrm{FT}} = C_{I\mathrm{,}\max}^{\mathrm{FT}} = I_{\max}^{\mathrm{FT}} + 1.
\end{equation}
Since the resolution process always begins with an initial collision, the number of collisions exceeds the number of idle intervals by exactly one, as shown in Figs. \ref{fig: validation of analytical tree bounds}b and \ref{fig: validation of analytical tree bounds}d.

\subsubsection{Last Success (LS) Scenario}
The LS scenario characterizes the resolution process up to the final successful transmission, excluding silent intervals occurring after all contending devices have transmitted. This distinction is relevant for access latency analysis, since such idle intervals do not contribute to the experienced latency.

Under the LS scenario, the maximum number of idle intervals is given by
\begin{equation}
I_{\max}^{\mathrm{LS}} =
2  \left\lfloor \frac{N-2}{5} \right\rfloor
+ 1_{\{q \geq 2\}}, \quad q = (N-2) \bmod 5.
\end{equation}

This expression exhibits a step-wise growth with increasing network size~$N$, shifted by two positions (Fig. \ref{fig: validation of analytical tree bounds}e). Within each period of five network sizes, two step increases occur: one determined by the floor term $\lfloor (N-2)/5 \rfloor$, and the other by the indicator term $1_{\{q \geq 2\}}$.

The worst-case behavior occurs when colliding pairs occupy the first $\lfloor (M-1)/2 \rfloor$ groups, $\mathcal{S}_{1}, \dots,\mathcal{S}_{\lfloor (M-1)/2 \rfloor}$, as well as the last group, $\mathcal{S}_{M}$. Similar to the FT scenario, collisions within the earlier groups maximize the number of idle intervals due to their larger set sizes. Including the final group delays the last successful transmission, thereby maximizing the accumulation of idle intervals before its occurrence.

The maximum number of collisions in the LS scenario remains
\begin{equation}
C_{\max}^{\mathrm{LS}} = C_{\max}^{\mathrm{FT}},
\end{equation}
since collisions cannot occur after the final successful transmission (Fig. \ref{fig: validation of analytical tree bounds}f).

Beyond these primary bounds, the conditional bounds are given as follows. The maximum number of idle intervals under maximum collisions is given by
\begin{equation}
I_{C\mathrm{,}\max}^{\mathrm{LS}} =
\begin{cases}
\left\lfloor \frac{N}{5} \right\rfloor-1, & p = 0, \\
\left\lfloor \frac{N}{5} \right\rfloor, & p = 1, \\
2 \cdot \left\lfloor \frac{N}{5} \right\rfloor, & p = 2,3, \\
2 \cdot \left\lfloor \frac{N}{5} \right\rfloor + 1, & p = 4,
\end{cases}
\end{equation}
where $p = N \bmod 5$. This expression exhibits a piecewise step behavior with increasing network size $N$ (Fig. \ref{fig: validation of analytical tree bounds}g). The term $\lfloor N/5 \rfloor$ determines the baseline growth of idle intervals, while the remainder $p$ governs the oscillatory behavior around this baseline.

Under maximum idle intervals, the maximum number of collisions is expressed as
\begin{equation}
C_{I,\max}^{\mathrm{LS}} =
\begin{cases}
I_{\max}^{\mathrm{LS}} + 2, & N = 3,6, \\
I_{\max}^{\mathrm{LS}} + 1, & \text{otherwise}.
\end{cases}
\end{equation}
In general, the maximum number of collisions exceeds the maximum number of idle intervals by one, except for $N=3$ and $N=6$, where an extra collision occurs whose corresponding idle interval lies after the final successful transmission (Fig.~\ref{fig: validation of analytical tree bounds}h).

\subsubsection{First Success (FS) Scenario}
The FS scenario characterizes the resolution process up to the first successful transmission. This scenario directly corresponds to the SP configuration, where the CTO counter is reset immediately after the first successful transmission. Consequently, collisions and idle intervals occurring afterward do not contribute to the experienced latency. 

Under the FS scenario, the maximum number of idle intervals is given by
\begin{equation}
I_{\max}^{\mathrm{FS}} =
( z - \left\lceil \frac{z^2}{N} \right\rceil )
 \left( \left\lceil \log_z(N) \right\rceil - 1 \right), \quad z = \lceil N/2 \rceil,
\label{equ: I max FS long}
\end{equation}
which can be simplified to
\begin{equation}
I_{\max}^{\mathrm{FS}} =
\left\lfloor \frac{N}{4} \right\rfloor.
\label{equ: I max FS short}
\end{equation}
Equations (\ref{equ: I max FS long}) and (\ref{equ: I max FS short}) are equivalent and exhibit a step-wise growth with increasing network size~$N$, increasing by one idle interval for each period of four network sizes (Fig.~\ref{fig: validation of analytical tree bounds}i).

In contrast to the FT and LS scenarios, where worst-case behavior arises when colliding pairs occupy the earliest (and typically larger) groups, the worst-case FS scenario occurs when the first successful transmission is delayed to the last group. This is achieved when the colliding devices possess the highest node IDs, i.e., $\{N-M+1,\dots,N\}$ and the number of colliding devices $M$ is sufficiently large to maximize traversal through idle branches before reaching the first successful transmission.

Multiple values of $M$ can produce this worst-case behavior. One representative case is $M = \lceil N/2 \rceil$, which applies for all $N > 2$, since at least two devices must collide to initiate the resolution process.

Furthermore, the maximum number of collisions in the FS scenario is given by
\begin{equation}
C_{\max}^{\mathrm{FS}} = \left\lceil \log_2(N) \right\rceil,
\end{equation}
which grows logarithmically with the network size~$N$ (Fig.~\ref{fig: validation of analytical tree bounds}j). The number of collisions corresponds to the height of the resolution tree, i.e., $\lceil \log_M(N) \rceil$, and is maximized when the branching factor is minimal ($M=2$), forcing the resolution process to traverse the largest possible number of levels.

Beyond these primary bounds, the conditional bounds are given as follows. Under maximum collisions, the maximum number of idle intervals is obtained as
\begin{equation}
I_{C\mathrm{,}\max}^{\mathrm{FS}} = \left\lfloor \log_2( N - 2^{C_{\max}^{\mathrm{FS}} - 1} ) \right\rfloor.
\end{equation}
This expression exhibits a piecewise step behavior with increasing network size~$N$ (Fig.~\ref{fig: validation of analytical tree bounds}k). Its upper envelope grows logarithmically with $N$, while the function oscillates between consecutive powers of two.

Finally, under maximum idle intervals, the maximum number of collisions is expressed as
\begin{equation}
C_{I\mathrm{,}\max}^{\mathrm{FS}} =
\begin{cases}
1, & N = 2,\\
3, & N \in \{6,7,8,9,10,11,15,19\},\\
2, & \text{otherwise},\quad N \geq 2.
\end{cases}
\end{equation}
The maximum number of collisions prior to the first successful transmission depends on the network size as specified above and is bounded by three (Fig.~\ref{fig: validation of analytical tree bounds}l).

\subsection{Resulting Worst-Case Latency Bounds}
\label{subsubsection: latency bounds}

Building on these resolution bounds, characterized in Section \ref{subsubsec:treelatency}, we derive the resulting worst-case latency bounds for the highest-priority traffic ($k=1$) in the SP configuration (PLCA-SP). Under contention, the access latency includes multiple COMMIT attempts, corresponding to collisions, as well as silent intervals incurred during the resolution process before the channel is successfully reserved. Since the silent interval duration $T_\mathrm{s}$ is significantly larger than the collision duration $T_\mathrm{c}$ (Table~\ref{tab:PLCA Parameters}), the worst-case latency is primarily dominated by the number of idle intervals. Accordingly, all bounds derived in the following assume the maximum number of idle intervals, $I_{\max}$, together with the corresponding maximum number of collisions under this condition, i.e.,~$C_{I,\max}$.

We further distinguish between node-based and class-based access latency. Node-based access latency, defined in Section~\ref{subsec:system model}, denotes the time elapsed between the instant a packet becomes ready for transmission at device $n$ and the instant its transmission begins. In contrast, class-based access latency denotes the time elapsed between the instant at least one packet of class $k$ becomes ready for transmission at any device and the instant the first packet of class $k$ begins transmission. 

The node-based definition applies directly to node-based access mechanisms, such as standard PLCA and PLCA-NP. It also applies to priority-based access mechanisms, including PLCA-SP, under the assumption that at most one packet of a given priority class is present in the system at any time. This single-packet assumption enables a fair comparison with PLCA-NP, where each priority class is associated with a single device. With this assumption, both PLCA-NP and PLCA-SP operate under equivalent conditions.

\subsubsection{Class-based Access Latency (Single-Packet Assumption)}

We first consider the single-packet assumption, where at most one packet of the highest-priority class is present in the system at any time. In PLCA-SP, the worst-case access latency for the highest-priority traffic is governed by the FS resolution scenario. Since the CTO counter is reset to CTO-$1$ immediately after the first successful transmission, collisions and silent intervals occurring afterward do not contribute to the experienced latency.

Accordingly, the worst-case access latency for the highest-priority traffic, under the single-packet assumption, is given by
\begin{multline}
L_{k=1\mathrm{,}\max}^{\text{PLCA-SP,~single}} =
T_\mathrm{b}
+ (K-2) T_\mathrm{s}
+ 2  T_c
+ T_{\mathrm{pkt,}\max} \\
+ I_{\max}^{FS}  T_\mathrm{s}
+ C_{I,\max}^{FS}  T_\mathrm{c}.
\label{equ: SP access latency single}
\end{multline}
The term ($I_{\max}^{\mathrm{FS}}T_\mathrm{s} + C_{I,\max}^{\mathrm{FS}}T_\mathrm{c}$) represents the maximum resolution delay associated with a \emph{foreign} contention tree, i.e., a resolution process corresponding to a traffic class $k \neq 1$.

In the worst-case scenario, a highest-priority packet becomes ready immediately after the occurrence of CTO-1. This packet must then wait for $(K-2)$ silent CTOs, followed by one active CTO (i.e., CTO-$K$) incurring the maximum resolution delay and a packet transmission of maximum duration~$T_{\mathrm{pkt,}\max}$.

This behavior is illustrated in Fig.~\ref{PNA-SP and Tree Transmission Sequence Example} for a network with $N=8$ devices and $K=3$ priority classes. The worst-case access latency for the highest-priority class, under the single-packet assumption, is bounded by $1.2285$~ms (Fig.~\ref{PNA-SP Transmission Sequence Example}). This is significantly lower than the $8.5068$~ms observed with standard PLCA and comparable to, yet slightly lower than, the $1.2457$~ms achieved by PLCA-NP. Unlike PLCA-NP, however, PLCA-SP allows the highest-priority packet to originate from any device, even under the single-packet assumption.

\input{Fig6_PLCA_SP}

\subsubsection{Class-based Access Latency (Multi-Packet Case)}
We next consider the general case, where multiple devices may simultaneously generate packets of the highest-priority class. In this setting, node-based access latency is no longer sufficient to characterize system behavior, and class-based access latency is therefore considered.

The worst-case access latency for the highest-priority traffic remains governed by the FS resolution scenario and is given by
\begin{multline}
\label{equ:SP_worst_case_latency}
L_{k=1\mathrm{,}\max}^{\text{PLCA-SP}} =
T_\mathrm{b}
+ (K-2)\ T_\mathrm{s}
+ 2  T_\mathrm{c}
+ T_{\mathrm{pkt,}\max} \\
+ 2  (I_{\max}^{FS}  T_\mathrm{s}
+ C_{I\mathrm{,}\max}^{FS}  T_\mathrm{c} ).
\end{multline}

In this expression, the resolution delay term ($I_{\max}^{FS} T_\mathrm{s} +
C_{I\mathrm{,}\max}^{FS} T_\mathrm{c}$) appears twice, reflecting two consecutive contention phases. The first occurrence corresponds to the resolution delay associated with a \emph{foreign} contention tree, i.e., contention originating from lower-priority traffic classes, as in~(\ref{equ: SP access latency single}). The second corresponds to the delay incurred before the first successful transmission within the \emph{native} contention tree associated with class $k=1$. In both phases, the worst-case delay is governed by the FS scenario, since the access latency is determined by the first successful transmission within the class.

For a network with $N=8$ devices and $K=3$ priority classes, the resulting worst-case access latency for the highest-priority class is bounded by $1.2364$~ms. This bound remains significantly lower than that of standard PLCA and marginally lower than that achieved by PLCA-NP, while allowing multiple devices to simultaneously generate the highest-priority traffic.

\subsubsection{Class-based Completion Latency}
In addition to access latency, we consider the completion (drain) latency, defined as the time elapsed between the instant at least one packet of class $k$ becomes ready for transmission and the instant the last pending packet of class $k$ completes transmission. This metric provides an upper bound on the latency experienced by any packet of class $k$ and characterizes the time required to fully drain all pending traffic belonging to the class.

The worst-case completion latency for the highest-priority traffic is influenced by both the FS and LS resolution scenarios. The FS scenario determines the resolution delay until the first successful transmission within a \emph{foreign} contention tree, whereas the LS scenario defines the delay within the \emph{native} contention tree until the last pending packet of class $k=1$ gains access to the channel.

\begin{figure*}[!t]  
    \centering 
    \includegraphics[width=\linewidth]{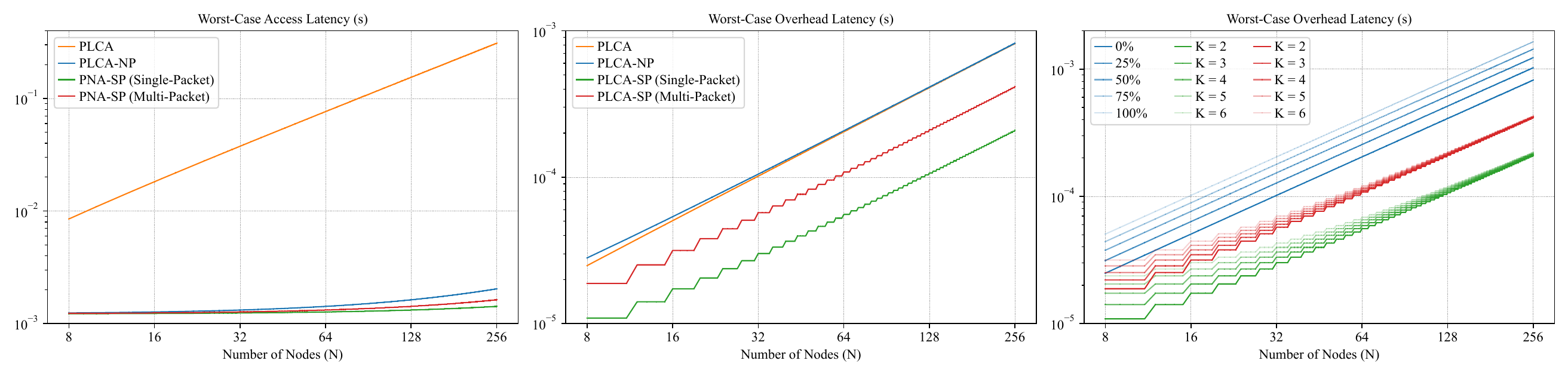}
    \caption{Theoretical worst-case access and overhead latency analysis of PLCA, PLCA-NP, PLCA-SP (single-packet), and PLCA-SP (multi-packet) as a function of network size ($N=8$ to $256$).}
    \label{fig: Theoretical latency and overhead analysis} 
\end{figure*}

Accordingly, the worst-case class completion latency is expressed as 

\begin{multline}
\label{equ:SP_worst_case_delay}
D_{k=1\mathrm{,}\max}^{\text{PLCA-SP}} =
L_{k=1\mathrm{,}\max}^{\text{PLCA-SP, single}}
+ I_{\max}^{\mathrm{LS}}  T_\mathrm{s}
+ C_{I\mathrm{,}\max}^{\mathrm{LS}}  T_\mathrm{c}  \\
+ \frac{\alpha_{k=1}}{C} 
+ (N - 3) T_\mathrm{c}
\end{multline}
where $\alpha_{k=1}$ denotes the maximum amount of pending class-$1$ traffic (in bits), and $C$ represents the channel data rate (in~bps). After the contention of at least two devices in the \emph{foreign} contention tree, up to $N-3$ additional COMMIT durations may still occur in the \emph{native} contention tree of class $k=1$, since the access latency expression in (\ref{equ: SP access latency single}) already includes one COMMIT message associated with class-$1$.

%% file: Fig6_PLCA_SP.tex
\begin{figure}[!t]

    \centering
    \subfloat[Worst-case resolution process during CTO-3 for contending devices $n=7,8$, illustrating the maximum number of idle intervals and the corresponding maximum number of collisions for $N=8$.]{ 
        \begin{tikzpicture}
        [
            level distance=1.5cm, 
            sibling distance=2cm 
        ]

        \node[draw=PriorityOL, fill=PriorityFill, circle, minimum size=0.8cm] {\tiny $7,8$}
            child[draw=WhiteOLL] {node[draw=WhiteOLD, fill=WhiteFillL, circle, minimum size=0.8cm] {\tiny $ $}}
            child [draw=WhiteOLL]{node[draw=PriorityOL, fill=PriorityFill, circle, minimum size=0.8cm] {\tiny $7,8$} 
                child [draw=WhiteOLL]{node[draw=WhiteOLD, fill=WhiteFillL, circle, minimum size=0.8cm] {\tiny $ $}} 
                child[draw=WhiteOLL]{node[draw=PriorityOL, fill=PriorityFill, circle, minimum size=0.8cm] {\tiny $7,8$}
                    child [draw=WhiteOLL]{node[draw=BlueOLD, fill=BlueFillD, circle, minimum size=0.8cm] {\tiny $7$}} 
                    child [draw=WhiteOLL]{node[draw=BlueOLD, fill=BlueFillD, circle, minimum size=0.8cm] {\tiny $8$}} 
                } 
            };
            
    \node[] at (0,0 + 0.6 + 0.25) {\tiny Set $\mathcal{N} =$};
    \node[] at (0,0 + 0.6) {\tiny $\{1,2,\dots,8\}$};

    \node[] at (0 - 1,-1.5 + 0.5 + 0.25) {\tiny Set $1 =$};
    \node[] at (0 - 1,-1.5 + 0.5) {\tiny $\{1,\dots,4\}$};

    \node[] at (0 + 1,-1.5 + 0.5 + 0.25) {\tiny Set $2 =$};
    \node[] at (0 + 1,-1.5 + 0.5) {\tiny $\{5,\dots,8\}$};
        
    \node[] at (0 + 1 -1,-1.5 -1.5 + 0.5 + 0.25) {\tiny Set $2.1 =$};
    \node[] at (0 + 1 -1,-1.5 -1.5 + 0.5) {\tiny $\{5,6\}$};

    \node[] at (0 + 1 +1,-1.5 -1.5 + 0.5 + 0.25) {\tiny Set $2.2 =$};
    \node[] at (0 + 1 +1,-1.5 -1.5 + 0.5) {\tiny $\{7,8\}$};

    \node[] at ((0 + 1 +1 -1,-1.5 -1.5 -1.5 +0.5 + 0.25) {\tiny Set $2.2.1 =$};
    \node[] at (0 + 1 +1 -1,-1.5 -1.5 -1.5 + 0.5) {\tiny $\{7\}$};

    \node[] at ((0 + 1 +1 +1,-1.5 -1.5 -1.5 + 0.5 + 0.25) {\tiny Set $2.2.2 =$};
    \node[] at (0 + 1 +1 +1,-1.5 -1.5 -1.5 + 0.5) {\tiny $\{8\}$};

    \node [] at (0 + 1 - 4,-1.5 -1.5 -1.5 - 0.25) { \color{white} \tiny .};
    \node [] at (0 + 1 + 4,-1.5 -1.5 -1.5 - 0.25) { \color{white} \tiny .};

        \end{tikzpicture}
        \label{Tree splitting mechanism for SP}
        }

   \subfloat[Transmission sequence and worst-case access latency in PLCA-SP for highest-priority traffic under the single-packet assumption.]{%
        \begin{tikzpicture}

        \draw[->] (0,0.8) -- (8.4,0.8) node[right]{};
        \node at (8.2, 0.8+0.15) {\tiny Time};

        \node (b20) [beacon] at (0.25,0.4) {};
        \node(p10) [priority2, right=0cm of b20]{\tiny 1};
        \node(p20) [priority2, right=0cm of p10] {\tiny 2};
        \node(p30) [priority2, right=0cm of p20] {\tiny 3};

        \node (s31) [silent, right=0cm of p30] {\tiny 3};
        \node(c32) [priority2, right=0cm of s31] {\tiny 3};
        \node (s33) [silent, right=0cm of c32] {\tiny 3};
        \node(c34) [priority2, right=0cm of s33] {\tiny 3};
        \node(c35) [commit, right=0cm of c34] {\tiny 7};
        \node (p36) [xxlarge_packet, right=0cm of c35] {\tiny 7};

        \node(c40) [priority2, right=0cm of p36] {\tiny 1};
        \node (p41) [min_packet, right=0cm of c40] {\tiny 3};

        \node(c50) [priority2, right=0cm of p41] {\tiny 1};
        \node(c51) [priority2, right=0cm of c50] {\tiny 2};
        \node(c52) [priority2, right=0cm of c51] {\tiny 3};
        \node (c53) [min_packet, right=0cm of c52] {\tiny 8};

        \node[above=-0.1cm of p10] () [CTO_silent] {\tiny 1};
        \node[above=-0.1cm of p20] () [CTO_silent] {\tiny 2};
        
        \node[above=-0.1cm of p30] () [CTO_commit] {\tiny 7};
        \node[above=0.1cm of p30] () [CTO_commit, xshift=0.2cm] {\tiny 8};

        \node[above=-0.1cm of c32] () [CTO_commit] {\tiny 7};
        \node[above=0.1cm of c32] () [CTO_commit, xshift=0.2cm] {\tiny 8};

        \node[above=-0.1cm of c34] () [CTO_commit] {\tiny 7};
        \node[above=0.1cm of c34] () [CTO_commit, xshift=0.2cm] {\tiny 8};
        
        \node[above=-0.1cm of c40] () [CTO_commit] {\tiny 3};

        \node[above=-0.1cm of c50] () [CTO_silent] {\tiny 1};
        \node[above=-0.1cm of c51] () [CTO_silent] {\tiny 2};

        \node[above=-0.1cm of c52] () [CTO_commit] {\tiny 8};

        \draw[->] (0.825,-0.4) -- (0.825,0); 
        \draw[->] (5.475,-0.4) -- (5.475,0);
        \draw[<->] (0.825 + 0.1,-0.2) -- (5.475 - 0.1,-0.2);
        \node[] at (0.825 + 0.01, -0.4 - 0.2) {\tiny Packet};
        \node[] at (0.825 + 0.02, -0.4 - 0.2 - 0.2) {\tiny is Ready};
        \node[] at (0.825 + 2.325, -0.2 - 0.2 - 0.1) {\tiny $L_{k=1\mathrm{,}\max}^{\text{PLCA-SP,~ single}} = 1.2285$ ms};
        \node[] at (5.475, -0.4 - 0.2) {\tiny Begin};
        \node[] at (5.475, -0.4 - 0.2 - 0.2) {\tiny Transmission};

        \begin{scope}[shift={(0, 1.9)}] 
        
        \node () [beaconStd] at (0.25+0.25,0) {};
        \node[right] at (0.25+0.15+0.25, 0) {\tiny BEACON};

        \node () [commitStd] at (2+0.25,0) {\tiny $n$};
        \node[right] at (2+0.15+0.25, 0) {\tiny COMMIT};
        
        \node () [ipg_priorityStd] at (3.75+0.25,0) {\tiny $k$};
        \node[right] at (3.75+0.15+0.25, 0) {\tiny Class};

        \node () [silentStd] at (5.375+0.25,0) {\tiny $k$};
        \node[right] at (5.375+0.15+0.25, 0) {\tiny Silent};

        \node () [dataStd] at (7+0.25,0) {\tiny $n$};
        \node[right] at (7+0.15+0.25, 0) {\tiny Data};
        \end{scope}
        
        \end{tikzpicture}
        \label{PNA-SP Transmission Sequence Example}
    }
    
    \caption{Operation and worst-case access latency of PLCA-SP in an $N=8$ device network with $K=3$ priority classes.}
    \label{PNA-SP and Tree Transmission Sequence Example}

\end{figure}

%% file: Ch7_evaluation.tex
\section{Evaluation} 
\label{Sec: Evaluation}

This section evaluates the proposed PLCA-SP mechanism in comparison with PLCA and PLCA-NP through both analytical and simulation-based analysis. First, the resolution latency bounds derived in Section~\ref{subsubsec:treelatency} are validated. Next, the scalability of the resulting worst-case access and overhead\footnote{Worst-case overhead latency refers to the maximum delay introduced by non-data transmission elements, including signaling overhead, such as BEACON and COMMIT messages, as well as contention resolution overhead such as collision and idle intervals prior to data transmission.} latencies is analyzed. Finally, the mechanisms are assessed through simulations conducted in OMNEST~\cite{omnest}.

\subsection{Validation of Resolution Latency Bounds}
\label{subsec: Validation of Resolution Latency Bounds}

We first validate the resolution latency bounds derived in Section~\ref{subsubsec:treelatency} and assess their tightness through exhaustive enumeration of contention scenarios. The analysis covers network sizes ranging from $N=2$ to $N=32$, with all combinations of contending devices exhaustively enumerated. For each network size and traversal scenario~$x \in \{\mathrm{FT},~\mathrm{LS},~\mathrm{FS}\}$, we record the numbers of idle intervals and collisions for every contention instance, together with the corresponding conditional quantities, and compare them with the bounds $I_{\max}^{x}$, $C_{\max}^{x}$, $I_{C,\max}^{x}$ and $C_{I,\max}^{x}$, respectively.

The observed results are shown in Fig.~\ref{fig: validation of analytical tree bounds}. Each column corresponds to a traversal scenario (FT, LS, FS), while each row represents one of the evaluated quantities. Solid lines denote the derived bounds, while scatter points represent the observed values. To improve visualization, the scatter points are slightly jittered, and darker regions indicate a higher concentration of observations.

As shown in Fig.~\ref{fig: validation of analytical tree bounds}, all derived bounds are valid across all evaluated network sizes and traversal scenarios. Each bound coincides with the corresponding observed maximum at every evaluated network size, confirming tightness over the evaluated range.

\subsection{Theoretical Performance and Scalability Analysis}

We next evaluate the worst-case access and overhead latencies of PLCA, PLCA-NP, and PLCA-SP as a function of network size. The evaluation of PLCA-NP and PLCA-SP focuses on the highest-priority traffic, while, for PLCA-SP, both the single-packet and multi-packet assumptions are considered. The single-packet assumption is primarily included to enable a fair comparison with PLCA-NP, where only one device is permitted to transmit the highest-priority traffic.

For PLCA-NP and PLCA-SP, the worst-case overhead latency occurs under the same conditions as their respective worst-case access latency. Therefore, the worst-case overhead is obtained by subtracting the packet transmission duration $T_{\mathrm{pkt,}\max}$ from the corresponding access latency expressions in (\ref{equ: PLCA-NP maximum latency}), (\ref{equ: SP access latency single}), and (\ref{equ:SP_worst_case_latency}).

PLCA, in contrast, exhibits distinct worst-case scenarios for access and overhead latencies. While worst-case access latency is derived in (\ref{equ: PLCA maximum latency}), the worst-case overhead corresponds to a different transmission sequence, illustrated in Cycle \#3 of Fig.~\ref{subfig: Transmission behavior of PLCA}, and is given by:  
\begin{equation}
    T_\mathrm{b} + (N-1) \cdot T_\mathrm{s} + T_\mathrm{c}
\end{equation}

Fig.~\ref{fig: Theoretical latency and overhead analysis} presents the worst-case access and overhead latencies as a function of the network size $N$, comparing PLCA, PLCA-NP, and PLCA-SP under both the single-packet and multi-packet assumptions. The first subplot shows the worst-case access latency, while the second subplot shows the corresponding worst-case overhead latency. For these two subplots, the number of priority classes is set to $K=2$. The third subplot illustrates the impact of increasing the number of priority nodes in PLCA-NP and the number of traffic classes $K$ in PLCA-SP.

The results, presented in the first subplot in Fig.~\ref{fig: Theoretical latency and overhead analysis}, show that PLCA-SP consistently achieves substantially lower worst-case access latency than standard PLCA across all network sizes. For example, in an 8-device network, PLCA-SP reduces the worst-case access latency by approximately 85.6\%, while in a 256-device network, the reduction reaches 99.4\%. Similar behavior is observed under the single-packet assumption.

Compared to PLCA-NP, PLCA-SP achieves comparable, or slightly lower, latency in small networks while providing greater flexibility, since high-priority traffic may originate from any device rather than being restricted to a single priority node. Moreover, the latency of PLCA-NP increases more significantly with network size. When scaling from 8 to 256 nodes, the access latency of PLCA-NP increases by approximately 63.9\%, whereas PLCA-SP exhibits a considerably smaller increase of only 16.2\% under the single-packet assumption and 32.1\% in the general multi-packet case.

\begin{figure*}[!t]  
    \centering 
    \includegraphics[width=\linewidth]{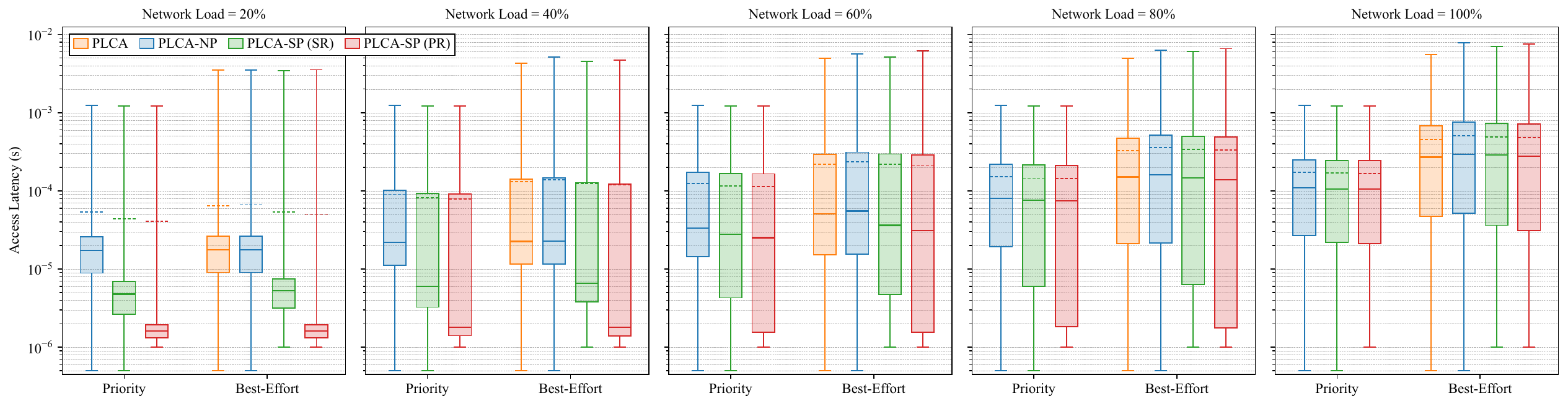}

    \caption{Access latency of PLCA, PLCA-NP, PLCA-SP (SR), and PLCA-SP (PR) in an eight-device network with one high-priority device under varying offered load ($20\%$--$100\%$). Boxes span the 25th--75th percentiles; solid and dashed lines indicate the median and mean, respectively; whiskers indicate the minimum and maximum observed values.}
    \label{fig: Access latency experienced by priority and best-effort nodes under varying network loads} 

\end{figure*}

\begin{figure*}[!t]  
    \centering 
    \includegraphics[width=\linewidth]{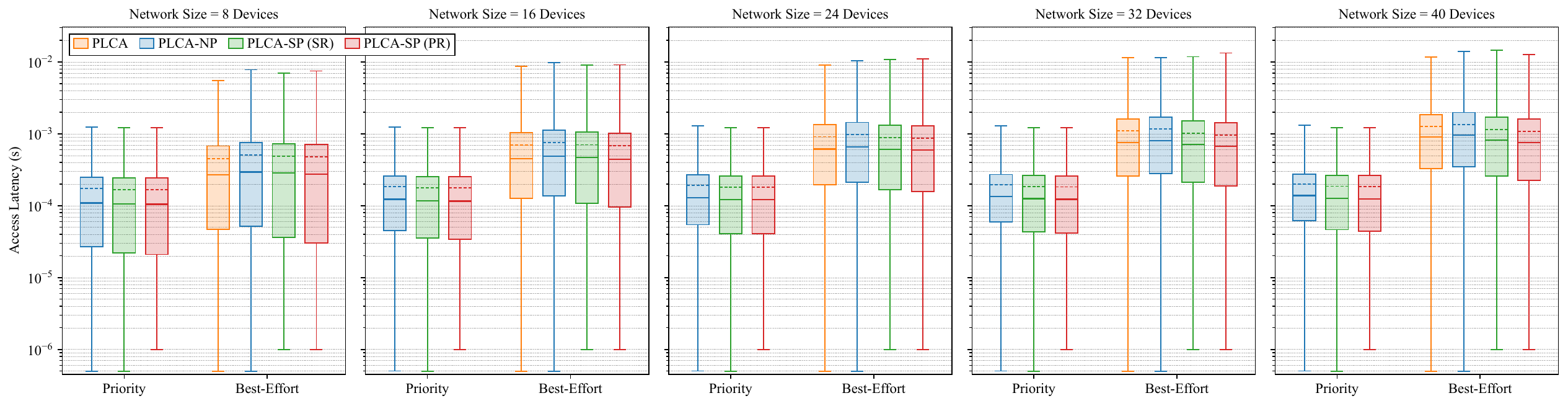}

    \caption{Access latency of PLCA, PLCA-NP, PLCA-SP (SR), and PLCA-SP (PR) with one high-priority device under varying network size ($N=8$--$40$). Boxes span the 25th--75th percentiles; solid and dashed lines indicate the median and mean, respectively; whiskers indicate the minimum and maximum observed values.}
    \label{fig: Access latency experienced by priority and best-effort nodes under varying network sizes} 

\end{figure*}

From an overhead perspective (second subplot in Fig.~\ref{fig: Theoretical latency and overhead analysis}), PLCA and PLCA-NP exhibit substantial overhead that increases linearly with network size due to their node-based cyclic access structure and the accumulation of idle intervals. In contrast, PLCA-SP achieves consistently lower overhead across all evaluated network sizes, reducing overhead latency by up to 74.6\% in large networks. 

Furthermore, as shown in the third subplot of Fig.~\ref{fig: Theoretical latency and overhead analysis}, the overhead of PLCA-NP increases with both network size and the number of priority nodes. When no priority nodes are configured, the overhead of PLCA-NP becomes equivalent to that of the standard PLCA. However, when all devices are configured as priority nodes, the overhead doubles. In contrast, PLCA-SP exhibits more controlled scaling behavior and maintains lower overhead than both PLCA and PLCA-NP even when multiple traffic classes are configured.

\subsection{Simulation-based Performance Evaluation}

Finally, we evaluate the performance of PLCA, PLCA-NP, and the proposed PLCA-SP mechanism through simulations. All mechanisms are implemented in OMNEST using the INET framework, and both PLCA-SP contention management strategies, i.e., SR and PR, are considered.

The simulated network consists of $N$ devices connected via a shared medium operating at 10~Mbps. Traffic is generated independently at each device according to a Poisson process, resulting in exponentially distributed inter-arrival times. Packet sizes follow a truncated exponential distribution with a mean of 250~B and are bounded between 64~B and 1518~B, in accordance with standard Ethernet frame size limits.

The evaluation considers two scenarios, with a single high-priority device and multiple high-priority devices, respectively, and examines performance under varying offered loads and network sizes. For each configuration, 25 independent simulation runs are performed, each using a different seed and a duration of 120~s. Latency statistics are calculated from packet observations pooled across all 25 runs. The latency distributions are presented as box plots, where boxes span the 25th to 75th percentiles, solid lines indicate the median, dashed lines indicate the mean, and whiskers extend to the minimum and maximum observed values.

\subsubsection{Single High-Priority Device}

In this scenario, one device generates high-priority traffic, while the remaining devices generate best-effort traffic. In PLCA-NP, the designated device operates as the priority node, while the remaining devices operate as regular nodes. In PLCA-SP, its traffic belongs to the highest-priority class, while traffic from the remaining devices belongs to the best-effort class. Standard PLCA serves as the baseline without traffic differentiation.

We first fix the network size to $N=8$ and vary the offered load from 20\% to 100\%. Fig.~\ref{fig: Access latency experienced by priority and best-effort nodes under varying network loads} presents the resulting access latency for high-priority and best-effort traffic. Both PLCA-SP strategies substantially reduce the maximum observed access latency for high-priority traffic compared to standard PLCA. At full load, the reduction reaches $77.9\%$, while at an offered load of $20\%$, it reaches $65.4\%$.

Compared to PLCA-NP, both PLCA-SP strategies achieve similar maximum latency for high-priority traffic while reducing mean and median latency, particularly at low offered loads. At an offered load of $20\%$, the mean latency is reduced by up to $23.8\%$.

For best-effort traffic, PLCA-SP achieves lower mean latency than PLCA-NP across all evaluated loads, but slightly higher than that of standard PLCA at high offered loads. At full load, the increase relative to standard PLCA is up to $7.8\%$, illustrating the trade-off between reducing high-priority latency and maintaining best-effort performance.

We next investigate scalability by varying the network size from $N=8$ to $N=40$. Fig.~\ref{fig: Access latency experienced by priority and best-effort nodes under varying network sizes} presents the resulting access latency. The results show favorable latency scalability for PLCA-SP compared with both PLCA and PLCA-NP. As the network size increases from 8 to 40 devices, the maximum observed high-priority latency under PLCA-NP increases by approximately $7.32\%$, whereas the corresponding increase under both PLCA-SP strategies remains below $1\%$.

In addition to improving latency for high-priority traffic, PLCA-SP maintains competitive latency for best-effort traffic. For large network sizes (i.e., $N>8$), both PLCA-SP strategies achieve equivalent or lower mean best-effort latency than standard PLCA. These results show that prioritizing high-priority traffic can also be accompanied by improved mean latency for best-effort traffic at larger network sizes.

\subsubsection{Multiple High-Priority Devices}

\begin{figure*}[!t]  
    \centering 
    \includegraphics[width=\linewidth]{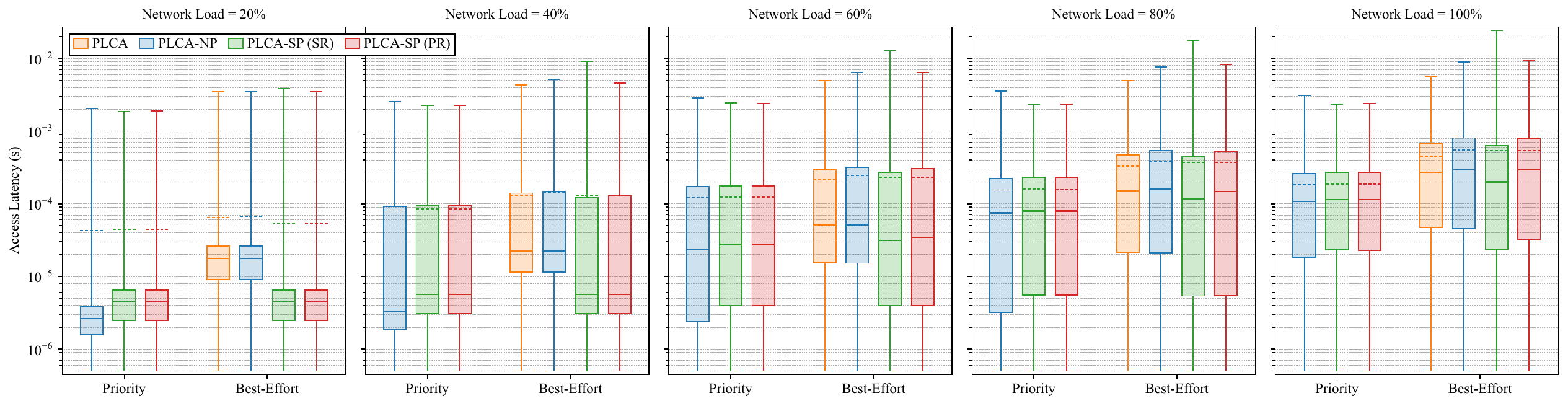}

    \caption{Access latency of PLCA, PLCA-NP, PLCA-SP (SR), and PLCA-SP (PR) in an eight-device network with two high-priority devices under varying offered load ($20\%$--$100\%$). Boxes span the 25th--75th percentiles; solid and dashed lines indicate the median and mean, respectively; whiskers indicate the minimum and maximum observed values.}
    \label{fig10: Access latency experienced by priority and best-effort nodes under varying network loads} 

\end{figure*}

\begin{figure}[!t]  
    \centering 
    \includegraphics[width=0.8\linewidth]{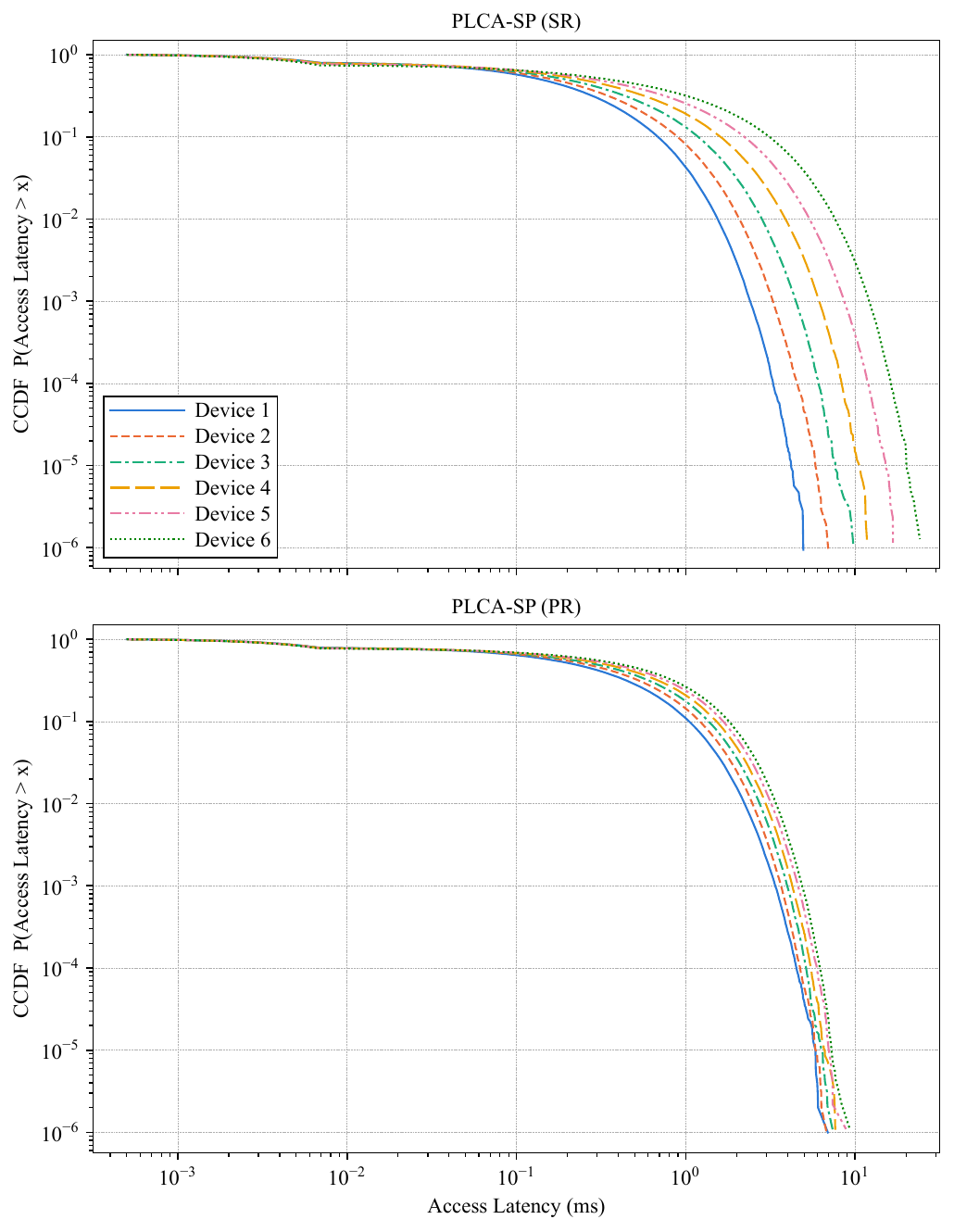}

    \caption{Complementary cumulative distribution functions (CCDFs) of access latency under PLCA-SP (SR) and PLCA-SP (PR) for the six best-effort devices in an eight-device network with two high-priority devices.}
    \label{fig11: CCDFs of SR and PR for BE  devices} 

\end{figure}

In this scenario, two devices generate high-priority traffic, while the remaining devices generate best-effort traffic. In PLCA-NP, the two designated devices operate as priority nodes with different priority ranks, while the remaining devices operate as regular nodes. In PLCA-SP, traffic from both designated devices belongs to the same highest-priority class, while traffic from the remaining devices belongs to the best-effort class. Standard PLCA again serves as the baseline without traffic differentiation. The network size is fixed to $N=8$, and the aggregate offered load varies from $20\%$ to $100\%$.

Fig.~\ref{fig10: Access latency experienced by priority and best-effort nodes under varying network loads} presents the access latency for high-priority and best-effort traffic under varying offered loads. Both PLCA-SP strategies achieve lower maximum observed high-priority latency than PLCA-NP across all evaluated loads,  with reductions of up to $34.1\%$. This demonstrates the benefit of assigning access opportunities to traffic classes rather than individual devices in reducing the maximum observed latency when multiple devices generate high-priority traffic.

We next examine the impact of the SR and PR strategies on best-effort traffic. In Fig.~\ref{fig10: Access latency experienced by priority and best-effort nodes under varying network loads}, their differences become more pronounced in the latency tail as the offered load increases. At full load, PR reduces the maximum observed best-effort latency by approximately $61.6\%$ relative to SR.

To investigate how these differences affect individual devices, Fig.~\ref{fig11: CCDFs of SR and PR for BE  devices} presents the complementary cumulative distribution functions (CCDFs) of access latency for the six best-effort devices, labeled devices~1--6, under SR and PR. SR exhibits substantial disparity among devices, with the ratio between the largest and smallest per-device 99th-percentile latencies reaching approximately $4.93$, compared with $1.54$ under PR. For device~6, PR reduces the 99th-percentile latency from approximately $7.62$~ms to $3.42$~ms, a reduction of $55.2\%$. This improved fairness among best-effort devices results from preserving contention-resolution progress across scheduling cycles. Notably, PR retains strict priority across classes; therefore, best-effort traffic can still experience starvation if higher-priority traffic remains continuously pending.

%% file: Ch8_conclusion.tex
\section{Conclusion}  
\label{Sec: Conclusion}

Although PLCA enables collision-free network access in shared-medium networks, it exhibits fundamental limitations. Network access is organized on a node basis without considering traffic priority, which can result in significant delays for time-sensitive traffic due to lower-priority transmissions. Existing PLCA extensions introduce prioritization through node differentiation rather than traffic differentiation. They modify either the transmission frequency or the access order of selected priority nodes to reduce their access latency. However, such approaches restrict priority traffic to specific devices and do not provide network-wide traffic differentiation.

To address these limitations, this paper proposed PLCA-SP, a priority-based network access and scheduling mechanism for shared-medium environments. In contrast to conventional PLCA-based systems, where access is determined by node order, PLCA-SP grants medium access according to traffic priority, allowing high-priority traffic generated by any device to access the medium ahead of lower-priority traffic. To resolve contention among packets of the same priority class, a deterministic ID-based resolution mechanism was introduced. Unlike conventional resolution schemes, branch selection is performed deterministically according to the device identifier, resulting in predictable behavior and bounded resolution delay. Furthermore, a configuration enabling SP scheduling over a shared medium was presented. The proposed mechanism was also extended with two contention management strategies, i.e., SR and PR, with different trade-offs between fairness and complexity.

The latency characteristics of the resolution process were analyzed under three traversal scenarios, i.e., FT, LS, and FS, from which analytical bounds were derived, validated through exhaustive evaluation, and shown to be tight. Building on these results, worst-case bounds were derived for both access and completion latency under the SP configuration. Analytical and simulation-based evaluations demonstrated that PLCA-SP significantly reduces worst-case access latency compared to standard PLCA, while achieving latency marginally lower than PLCA-NP. In contrast to PLCA-NP, however, PLCA-SP allows high-priority traffic to originate from any device in the network. Furthermore, the evaluations showed better scalability for PLCA-SP with increasing network size. While the latency of PLCA and PLCA-NP grows with the number of network devices, PLCA-SP maintains near-constant latency behavior.

The proposed mechanism introduces class-based transmission opportunities that allow simultaneous channel-reservation attempts and potential contention, requiring access-control functionality beyond standard PLCA. Future work will investigate the practical realization of these additional functions, assess compatibility with existing physical-layer implementations, and determine required hardware modifications.